\documentclass[10pt]{article}
\usepackage{axodraw}

\makeatletter
\@addtoreset{equation}{section}
\@addtoreset{figure}{section}
\makeatother
\newcommand{\be}{\begin{equation}}
\newcommand{\ee}{\end{equation}}
\newcommand{\ba}{\begin{eqnarray}}
\newcommand{\ea}{\end{eqnarray}}

\newcommand{\eq}{Eq.~}
\newcommand{\eqs}{Eqs.~}

\newcommand{\re}{Ref.~}
\newcommand{\nr}[1]{(\ref{#1})}
\newcommand{\nn}{\nonumber \\}
\newcommand{\fr}[2]{{\frac{#1}{#2}\,}}
\renewcommand{\(}{\left(}
\renewcommand{\)}{\right)}
\newcommand{\lb}{\left\{}
\newcommand{\rb}{\right\}}
\newcommand{\lk}{\left[}
\newcommand{\rk}{\right]}
\newcommand{\ld}{\left.}
\newcommand{\rd}{\right.}
\newcommand{\e}{\epsilon}
\newcommand{\ep}{\;\e}

\newcommand{\order}[1]{{\cal O}\(#1\)\vphantom{\fr12}}
\newcommand{\nN}{\nn[-3pt]}
\newcommand{\li}{{\rm Li}}
\newcommand{\ls}{{\rm Ls}}
\def\Asc(#1,#2)(#3,#4,#5){\CArc(#1,#2)(#3,#4,#5)}
\def\Lsc(#1,#2)(#3,#4){\Line(#1,#2)(#3,#4)}
\def\Ahh(#1,#2)(#3,#4,#5){\DashCArc(#1,#2)(#3,#4,#5){2}}
\def\Lhh(#1,#2)(#3,#4){\DashLine(#1,#2)(#3,#4){2}}
\def\scfc{0.7}  % picture scale factor
\def\phgt{21}   % all picture height 30 * \scfc
\def\pwc{21}    % c   picture width  30 * \scfc
\def\pwcb{31.5} % cb  picture width  45 * \scfc
\newcommand{\PIC}[4]{\;\parbox[c]{#2 pt}{\begin{picture}(#2,#3)(0,0)
\SetWidth{1.0}\SetScale{#4} #1 \end{picture}}\;}
\newcommand{\pic}[1]{\PIC{#1}{\pwc}{\phgt}{\scfc}}
\newcommand{\picb}[1]{\PIC{#1}{\pwcb}{\phgt}{\scfc}}
\def\TopoVR(#1){\pic{#1(15,15)(15,-90,270)}}
\def\TopoVRtad(#1,#2){\pic{#1(15,15)(15,-90,270) #2(5,0)(25,0)}}
\def\ToptVS(#1,#2,#3){\pic{#1(15,15)(15,0,180) #2(15,15)(15,180,360)%
 #3(30,15)(0,15)}}
\def\ToprVM(#1,#2,#3,#4,#5,#6){\pic{#3(15,15)(15,-30,90) #1(15,15)(15,90,210)%
 #2(15,15)(15,210,330) #5(2,7.5)(15,15) #6(15,15)(15,30) #4(28,7.5)(15,15)}}
\def\ToprVV(#1,#2,#3,#4,#5){\!\!\picb{#2(26.25,15)(15,256,76)%
 #3(30,30)(15,30) #1(18.75,15)(15,104,284) #4(15,30)(22.5,0)%
 #5(30,30)(22.5,0)}\!\!}
\def\ToprVB(#1,#2,#3,#4){\picb{#1(30,15)(15,-120,120) #2(30,15)(15,120,240)%
 #3(15,15)(15,60,300) #4(15,15)(15,-60,60)}}
\def\TopfVX(#1,#2,#3,#4,#5,#6,#7,#8,#9){\picb{#1(15,15)(15,90,270)%
 #2(30,15)(15,-90,90) #4(30,30)(15,30) #3(15,0)(30,0) #6(15,0)(15,15)%
 #5(15,15)(30,30) #8(15,30)(20,25) #8(25,20)(30,15) #7(30,15)(30,0)%
 #9(15,15)(30,15)}}
\def\TopfVH(#1,#2,#3,#4,#5,#6,#7,#8,#9){\picb{#1(15,15)(15,90,270)%
 #2(30,15)(15,-90,90) #4(30,30)(15,30) #3(15,0)(30,0) #6(15,0)(15,15)%
 #5(15,15)(15,30) #8(30,30)(30,15) #7(30,15)(30,0) #9(15,15)(30,15)}}
\def\TopfVW(#1,#2,#3,#4,#5,#6,#7,#8){\pic{#1(15,15)(15,90,180)%
 #3(15,15)(15,180,270) #2(15,15)(15,270,360) #4(15,15)(15,0,90)%
 #5(15,15)(15,30) #7(15,15)(15,0) #6(0,15)(15,15) #8(30,15)(15,15)}}
\def\TopfVWdot(#1,#2,#3,#4,#5,#6,#7,#8){\pic{#1(15,15)(15,90,180)%
 #3(15,15)(15,180,270) #2(15,15)(15,270,360) #4(15,15)(15,0,90)%
 #5(15,15)(15,30) #7(15,15)(15,0) #6(0,15)(15,15) #8(30,15)(15,15)%
 \Vertex(22,15){2}}}
\def\TopfVWlap(#1,#2,#3,#4,#5,#6,#7,#8){\pic{#1(15,15)(15,90,180)%
 #3(15,15)(15,180,270) #2(15,15)(15,270,360) #4(15,15)(15,0,90)%
 #5(15,15)(15,30) #7(15,15)(15,0) #6(0,15)(15,15) #8(30,15)(15,15)%
 \Text(2,19)[br]{$\scriptstyle 1$}\Text(20,2)[tl]{$\scriptstyle 2$}}}
\def\TopfVV(#1,#2,#3,#4,#5,#6,#7,#8){\!\!\picb{#2(26.25,15)(15,256,346)%
 #3(26.25,15)(15,-14,76) #4(30,30)(15,30) #1(18.75,15)(15,104,284)%
 #7(22.5,0)(15,30) #6(30,30)(26.25,15) #8(26.25,15)(22.5,0)%
 #5(25.25,15)(39.8,11.4)}\!\!}
\def\TopfVB(#1,#2,#3,#4,#5,#6,#7){\picb{#2(30,15)(15,-120,120)%
 #6(30,15)(15,120,180) #5(30,15)(15,180,240) #1(15,15)(15,60,300)%
 #4(15,15)(15,-60,0) #3(15,15)(15,0,60) #7(30,15)(15,15)}}
\def\TopfVBdot(#1,#2,#3,#4,#5,#6,#7){\picb{#2(30,15)(15,-120,120)%
 #6(30,15)(15,120,180) #5(30,15)(15,180,240) #1(15,15)(15,60,300)%
 #4(15,15)(15,-60,0) #3(15,15)(15,0,60) #7(30,15)(15,15)%
 \Vertex(28,22){2}}}
\def\TopfVBlap(#1,#2,#3,#4,#5,#6,#7){\picb{#2(30,15)(15,-120,120)%
 #6(30,15)(15,120,180) #5(30,15)(15,180,240) #1(15,15)(15,60,300)%
 #4(15,15)(15,-60,0) #3(15,15)(15,0,60) #7(30,15)(15,15)%
 \Text(9,15)[r]{$\scriptstyle 1$}\Text(22.5,15)[l]{$\scriptstyle 2$}}}
\def\TopfVN(#1,#2,#3,#4,#5,#6,#7){\picb{#1(15,15)(15,90,270)%
 #2(30,15)(15,-90,90) #4(30,30)(15,30) #3(15,0)(30,0)%
 #5(15,0)(15,30) #6(30,30)(30,0) #7(15,30)(30,0)}}
\def\TopfVU(#1,#2,#3,#4,#5,#6,#7){\pic{#3(15,15)(15,0,90)%
 #2(15,15)(15,90,180) #4(15,15)(15,180,270) #1(15,15)(15,270,360)%
 #6(0,15)(15,30) #7(15,0)(0,15) #5(30,15)(15,0)}}
\def\TopfVT(#1,#2,#3,#4,#5,#6){\pic{#1(15,15)(15,90,210)%
 #2(15,15)(15,210,330) #3(15,15)(15,-30,90) #4(2,7.5)(15,30)%
 #6(28,7.5)(2,7.5) #5(15,30)(28,7.5)}}
\def\TopLV(#1,#2,#3,#4,#5,#6){\!\!\picb{#2(26.25,15)(15.5,256,76)%
 #3(30,30)(15,30) #1(18.75,15)(15.5,104,284) #4(15,30)(22.5,0)%
 #5(30,30)(22.5,0) #6(15,17.8)(19.3,292.8,39.1)}\!\!}
\def\TopfVBB(#1,#2,#3,#4,#5){\picb{#1(30,15)(15,-120,120)%
 #2(30,15)(15,120,240) #3(15,15)(15,60,300) #4(15,15)(15,-60,60)%
 #5(22.5,3)(22.5,27)}}
\def\TopfVBBdot(#1,#2,#3,#4,#5){\picb{#1(30,15)(15,-120,120)%
 #2(30,15)(15,120,240) #3(15,15)(15,60,300) #4(15,15)(15,-60,60)%
 #5(22.5,3)(22.5,27) \Vertex(22.5,10){2} \Vertex(22.5,20){2}}}
\def\TopfVBBlap(#1,#2,#3,#4,#5){\picb{#1(30,15)(15,-120,120)%
 #2(30,15)(15,120,240) #3(15,15)(15,60,300) #4(15,15)(15,-60,60)%
 #5(22.5,3)(22.5,27)%
 \Text(2,10.5)[l]{$\scriptstyle 1$}\Text(29.5,10.5)[r]{$\scriptstyle 2$}}}

\def\oneJ{\TopoVRtad(\Asc,\Lsc)}
\def\twoS{\ToptVS(\Asc,\Asc,\Lsc)}
\def\twoSa{\ToptVS(\Asc,\Asc,\Lhh)}
\def\twoSb{\ToptVS(\Ahh,\Ahh,\Lsc)}
\def\threeB{\ToprVB(\Asc,\Asc,\Asc,\Asc)}
\def\threeBa{\ToprVB(\Ahh,\Ahh,\Asc,\Asc)}
\def\threeBb{\ToprVB(\Ahh,\Asc,\Asc,\Asc)}
\def\threeBc{\ToprVB(\Ahh,\Ahh,\Asc,\Ahh)}
\def\threeV{\ToprVV(\Asc,\Asc,\Lsc,\Lsc,\Lsc)}
\def\threeVa{\ToprVV(\Asc,\Asc,\Lsc,\Lhh,\Lsc)}
\def\threeVb{\ToprVV(\Ahh,\Asc,\Lsc,\Lhh,\Lsc)}
\def\threeVc{\ToprVV(\Asc,\Asc,\Lsc,\Lhh,\Lhh)}
\def\threeVd{\ToprVV(\Asc,\Ahh,\Lsc,\Lhh,\Lhh)}
\def\threeVe{\ToprVV(\Ahh,\Ahh,\Lsc,\Lhh,\Lhh)}
\def\threeVf{\ToprVV(\Asc,\Asc,\Lhh,\Lsc,\Lsc)}
\def\threeVg{\ToprVV(\Asc,\Asc,\Lhh,\Lsc,\Lhh)}
\def\threeVh{\ToprVV(\Asc,\Ahh,\Lhh,\Lsc,\Lhh)}
\def\threeVi{\ToprVV(\Asc,\Asc,\Lhh,\Lhh,\Lhh)}
\def\threeVj{\ToprVV(\Asc,\Ahh,\Lhh,\Lhh,\Lhh)}
\def\threeM{\ToprVM(\Asc,\Asc,\Asc,\Lsc,\Lsc,\Lsc)}
\def\threeMa{\ToprVM(\Asc,\Asc,\Asc,\Lsc,\Lsc,\Lhh)}
\def\threeMb{\ToprVM(\Asc,\Ahh,\Asc,\Lsc,\Lsc,\Lhh)}
\def\threeMc{\ToprVM(\Asc,\Ahh,\Ahh,\Lsc,\Lsc,\Lsc)}
\def\threeMd{\ToprVM(\Asc,\Asc,\Asc,\Lhh,\Lhh,\Lhh)}
\def\threeMe{\ToprVM(\Asc,\Ahh,\Asc,\Lsc,\Lhh,\Lhh)}
\def\threeMf{\ToprVM(\Ahh,\Ahh,\Ahh,\Lsc,\Lsc,\Lsc)}
\def\threeMg{\ToprVM(\Ahh,\Ahh,\Ahh,\Lsc,\Lsc,\Lhh)}
\def\threeMh{\ToprVM(\Ahh,\Asc,\Ahh,\Lhh,\Lhh,\Lsc)}
\def\threeMi{\ToprVM(\Ahh,\Ahh,\Ahh,\Lhh,\Lhh,\Lsc)}
\def\fourBBa{\TopfVBB(\Asc,\Ahh,\Asc,\Ahh,\Lhh)}
\def\fourBBb{\TopfVBB(\Asc,\Asc,\Asc,\Asc,\Lhh)}

\def\fourGa{\TopLV(\Asc,\Asc,\Lsc,\Lhh,\Lhh,\Ahh)}
\def\fourGb{\TopLV(\Asc,\Asc,\Lsc,\Lhh,\Lsc,\Asc)}
\def\fourGc{\TopLV(\Asc,\Asc,\Lhh,\Lsc,\Lhh,\Asc)}
\def\fourGd{\TopLV(\Asc,\Ahh,\Lhh,\Lsc,\Lhh,\Ahh)}
\def\fourGe{\TopLV(\Ahh,\Asc,\Lhh,\Lhh,\Lhh,\Asc)}
\def\fourT{\TopfVT(\Asc,\Asc,\Asc,\Lsc,\Lsc,\Lsc)}
\def\fourTa{\TopfVT(\Ahh,\Asc,\Ahh,\Lhh,\Lhh,\Lsc)}
\def\fourTb{\TopfVT(\Ahh,\Ahh,\Ahh,\Lsc,\Lsc,\Lsc)}
\def\fourTc{\TopfVT(\Ahh,\Asc,\Asc,\Lhh,\Lsc,\Lsc)}

\def\fourVBa{\TopfVB(\Ahh,\Asc,\Asc,\Ahh,\Asc,\Ahh,\Lsc)}
\def\fourVBb{\TopfVB(\Asc,\Asc,\Asc,\Ahh,\Ahh,\Asc,\Lsc)}
\def\fourVBc{\TopfVB(\Ahh,\Ahh,\Asc,\Ahh,\Ahh,\Asc,\Lsc)}
\def\fourVBd{\TopfVB(\Asc,\Asc,\Ahh,\Ahh,\Ahh,\Ahh,\Lhh)}
\def\fourVBe{\TopfVB(\Asc,\Ahh,\Asc,\Asc,\Ahh,\Ahh,\Lhh)}
\def\fourVBf{\TopfVB(\Ahh,\Ahh,\Asc,\Asc,\Asc,\Asc,\Lhh)}
\def\fourVBg{\TopfVB(\Asc,\Asc,\Asc,\Asc,\Asc,\Asc,\Lhh)}

\def\fourNa{\TopfVN(\Asc,\Asc,\Lsc,\Lsc,\Lhh,\Lhh,\Lhh)}
\def\fourNb{\TopfVN(\Asc,\Ahh,\Lsc,\Lhh,\Lhh,\Lhh,\Lsc)}
\def\fourNc{\TopfVN(\Asc,\Ahh,\Lhh,\Lhh,\Lsc,\Lhh,\Lhh)}
\def\fourNd{\TopfVN(\Asc,\Asc,\Lhh,\Lhh,\Lsc,\Lsc,\Lhh)}
\def\fourNe{\TopfVN(\Asc,\Asc,\Lsc,\Lhh,\Lhh,\Lsc,\Lsc)}

\def\fourUa{\TopfVU(\Asc,\Asc,\Asc,\Asc,\Lhh,\Lhh,\Lhh)}
\def\fourUb{\TopfVU(\Asc,\Asc,\Ahh,\Asc,\Lsc,\Lsc,\Lsc)}
\def\fourUc{\TopfVU(\Ahh,\Asc,\Ahh,\Asc,\Lhh,\Lsc,\Lsc)}
\def\fourUd{\TopfVU(\Ahh,\Ahh,\Ahh,\Asc,\Lhh,\Lhh,\Lsc)}

\def\fourVVa{\TopfVV(\Asc,\Ahh,\Ahh,\Lhh,\Lhh,\Lhh,\Lsc,\Lhh)}
\def\fourVVb{\TopfVV(\Ahh,\Asc,\Ahh,\Lhh,\Lsc,\Lhh,\Lhh,\Lsc)}
\def\fourVVc{\TopfVV(\Ahh,\Asc,\Asc,\Lhh,\Lhh,\Lsc,\Lhh,\Lsc)}
\def\fourVVd{\TopfVV(\Asc,\Asc,\Asc,\Lsc,\Lhh,\Lhh,\Lhh,\Lhh)}
\def\fourVVe{\TopfVV(\Asc,\Asc,\Ahh,\Lhh,\Lsc,\Lhh,\Lsc,\Lsc)}
\def\fourVVf{\TopfVV(\Asc,\Asc,\Ahh,\Lsc,\Lsc,\Lsc,\Lhh,\Lhh)}
\def\fourVVg{\TopfVV(\Asc,\Asc,\Asc,\Lhh,\Lhh,\Lsc,\Lsc,\Lsc)}

\def\fourWa{\TopfVW(\Asc,\Asc,\Asc,\Asc,\Lhh,\Lhh,\Lhh,\Lhh)}
\def\fourWb{\TopfVW(\Asc,\Asc,\Asc,\Ahh,\Lsc,\Lhh,\Lhh,\Lsc)}
\def\fourWc{\TopfVW(\Asc,\Asc,\Ahh,\Ahh,\Lsc,\Lsc,\Lsc,\Lsc)}
\def\fourWd{\TopfVW(\Ahh,\Asc,\Ahh,\Asc,\Lsc,\Lhh,\Lsc,\Lhh)}
\def\fourWe{\TopfVW(\Ahh,\Ahh,\Ahh,\Asc,\Lsc,\Lhh,\Lhh,\Lsc)}

\def\fourHa{\TopfVH(\Asc,\Asc,\Lhh,\Lhh,\Lsc,\Lsc,\Lsc,\Lsc,\Lhh)}
\def\fourHb{\TopfVH(\Asc,\Ahh,\Lhh,\Lhh,\Lsc,\Lsc,\Lhh,\Lhh,\Lhh)}
\def\fourHc{\TopfVH(\Asc,\Asc,\Lsc,\Lsc,\Lhh,\Lhh,\Lhh,\Lhh,\Lhh)}
\def\fourHd{\TopfVH(\Asc,\Asc,\Lhh,\Lsc,\Lhh,\Lsc,\Lsc,\Lhh,\Lsc)}
\def\fourHe{\TopfVH(\Ahh,\Asc,\Lhh,\Lsc,\Lsc,\Lhh,\Lsc,\Lhh,\Lsc)}

\def\fourXa{\TopfVX(\Asc,\Asc,\Lhh,\Lsc,\Lhh,\Lsc,\Lsc,\Lhh,\Lsc)}
\def\fourXb{\TopfVX(\Asc,\Asc,\Lsc,\Lsc,\Lhh,\Lhh,\Lhh,\Lhh,\Lhh)}
\def\ToprVMxxxx(#1,#2,#3,#4,#5,#6){\pic{#3(15,15)(15,-30,90)%
 #1(15,15)(15,90,210) \Vertex(15,22.5){2} \Text(15,15)[c]{$\scriptstyle x$}%
 #2(15,15)(15,210,330) #5(2,7.5)(15,15) #6(15,15)(15,30) #4(28,7.5)(15,15)}}
\def\threeMhx{\ToprVMxxxx(\Ahh,\Asc,\Ahh,\Lhh,\Lhh,\Lsc)}
\def\Mh{{M_2}}

\date{}

\begin{document}

%%%%%%%%%%%%%%%%%%%%%%%%%%%%%%%%%%%%%%%%%%%%%%%%%%%%%%%%%%%%
%%%%%%%%%%%%%%%%%%%%%%%%%%%%%%%%%%%%%%%%%%%%%%%%%%%%%%%%%%%%
%%%%%%%%%%%%%%%%%%%%%%%%%%%%%%%%%%%%%%%%%%%%%%%%%%%%%%%%%%%%

\title{
\centerline{\normalsize \mbox{} \hfill BI-TP 2005/11}
\vskip-1ex
\centerline{\normalsize \mbox{} \hfill UW/PT 05-5}
\vskip-1ex
\centerline{\normalsize \mbox{} \hfill hep-ph/0503209}
\vskip2ex
High-precision epsilon expansions of single-mass-scale 
four-loop vacuum bubbles
}
\author{\small Y.~Schr\"oder$^a$, A.~Vuorinen$^b$\\
{\small\it $^a$ Fakult\"at f\"ur Physik, Universit\"at Bielefeld, 
33501 Bielefeld, Germany}\\
{\small\it $^b$ Department of Physics, P.O. Box 351560, 
University of Washington, Seattle, WA 98195}}

\maketitle

\begin{abstract}
In this article we present a high-precision evaluation of the expansions
in $\e=(4-d)/2$ of (up to) four-loop scalar vacuum master integrals,
using the method of difference equations developed by S.~Laporta.
We cover the complete set of `QED-type' master integrals, i.e. those with
a single mass scale only (i.e. $m_i\in\{0,m\}$)
and an even number of massive lines at each vertex.
Furthermore, we collect all that is known analytically about 
four-loop `QED-type' masters, 
as well as about {\em all} single-mass-scale vacuum 
integrals at one-, two- and three-loop order.
\end{abstract}

%%%%%%%%%%%%%%%%%%%%%%%%%%%%%%%%%%%%%%%%%%%%%%%%%%%%%%%%%%%%
%%%%%%%%%%%%%%%%%%%%%%%%%%%%%%%%%%%%%%%%%%%%%%%%%%%%%%%%%%%%
%%%%%%%%%%%%%%%%%%%%%%%%%%%%%%%%%%%%%%%%%%%%%%%%%%%%%%%%%%%%

\section{Introduction}

Higher-order perturbative computations have become a necessity
in many areas of theoretical physics, be it
for high-precision tests of QED, QCD and the standard model,
or for studying critical phenomena in condensed matter systems.

Most recent investigations employ a highly automated approach,
utilizing algorithms that can be implemented on computer
algebra systems, in order to handle the growing numbers
of diagrams as well as integrals which occur at higher loop
orders.

Computations can be divided into four key steps.
First, the complete set of diagrams including symmetry factors
has to be generated.
For a detailed description of an algorithm for this step
for the case of vacuum topologies, see \re\cite{Kajantie:2001hv}.
Second, after specifying the Feynman rules, the color- and
Lorentz-algebra has to be worked out.
Third, within dimensional regularization,
massive use of the integration-by-parts (IBP) technique \cite{Chetyrkin:qh}
to derive linear relations between different Feynman integrals
in conjunction with an ordering prescription \cite{Laporta:2001dd} can 
be used to reduce the (typically large number of) integrals to a basis
of (typically a few) master integrals.
Practical notes as well as a classification of vacuum master
integrals are given in \re\cite{Schroder:2002re}.
Fourth, the master integrals have to be solved, either fully
analytically, or in an expansion around the space-time dimension $d$
of interest.
It is the fourth step that we wish to address here.

A very important subset of master integrals are fully massive
vacuum (bubble) integrals, since they constitute a main building
block in asymptotic expansions (see e.g. \re\cite{asyExp}).
They are also useful for massless
theories, when a propagator mass is introduced as an intermediate
infrared regulator \cite{Schroder:2003uw}.
In four dimensions, this class of master integrals has been
given up to the 4-loop level in \re\cite{Laporta:2002pg}.
As an application, these integrals are vital for computing the
4-loop QCD beta-function and anomalous dimensions \cite{betafct}.
In lower dimensions, perturbative results are needed
for applications in condensed matter systems, as well
as in the framework of dimensionally reduced effective
field theories for thermal QCD, where recent efforts have
made four-loop contributions an issue \cite{Kajantie:2002wa}.
We have recently extended the work of \re\cite{Laporta:2002pg},
to give the complete set of fully massive vacuum master integrals
in three dimensions, again up to the 4-loop level \cite{Schroder:2003kb}.

The next larger set of scalar vacuum master integrals are
those in which there is only one mass-scale $m$, i.e. the
propagators $1/(p_i^2+m_i^2)$ have masses $m_i\in\{0,m\}$.
These integrals are needed for problems with widely separated
mass scales, in which one then sets the masses of all heavy particles 
to $m$ and those of all light particles to zero. 
As a well-defined subset of these single-mass-scale integrals,
we here treat `QED-type' vacuum integrals,
i.e. those with an even number of massive lines at each vertex,
at the 4-loop level.
A recent application is in the computation of heavy-quark vacuum 
polarization \cite{Chetyrkin:2004fq}.

The complete set of `QED-type' vacuum master integrals up to 
the 4-loop level has already been identified in \re\cite{Schroder:2002re}.
The main purpose of this work is to numerically
compute this set in terms of a high-precision $\e$\/-expansion
in $d=4-2\e$ dimensions, and to present new analytic results 
for some low-order (in $\e$) coefficients.
Furthermore, we have made an attempt to collect all 
presently known analytic results on 4d single-mass-scale vacuum
integrals, up to four loops, in a coherent notation.

The plan of the paper is as follows.
In Section~\ref{se:two}, we give a brief review of the method
of difference equations applied to vacuum integrals.
In Section~\ref{se:three}, we discuss the actual implementation
of the algorithm.
In Section~\ref{se:four}, we display our numerical results for
the truncated power series expansions in $\e$ of
our master integrals, up to the four-loop level,
in $d=4-2\e$.
In Section~\ref{se:Laplace}, we discuss one case of a master
integral which we needed to solve via a Laplace transform of
its difference equation.
In Section~\ref{se:Analytic}, we list analytic results.

%%%%%%%%%%%%%%%%%%%%%%%%%%%%%%%%%%%%%%%%%%%%%%%%%%%%%%%%%%%%
%%%%%%%%%%%%%%%%%%%%%%%%%%%%%%%%%%%%%%%%%%%%%%%%%%%%%%%%%%%%
%%%%%%%%%%%%%%%%%%%%%%%%%%%%%%%%%%%%%%%%%%%%%%%%%%%%%%%%%%%%

\section{\label{se:two}
The evaluation of master integrals through difference equations}

The method we have chosen to compute the coefficients of the
truncated power series expansions of the master integrals is based
on constructing difference equations for the integrals and then
solving them numerically using factorial series. This approach has
recently been developed in \re\cite{Laporta:2001dd}, and below we
briefly summarize its basic concepts following the notation of the
original paper, which contains a much more detailed presentation of
the subject. While the method is completely general as it applies to
arbitrary kinematics, masses and topologies \cite{Laporta:2001rc},
our brief summary is somewhat adapted to the specific case of
vacuum integrals.

The main idea is to attach an arbitrary power $x$ to one of the
massive\footnote{The massiveness is crucial in order to avoid problems 
with the infrared behavior of the integral.} lines of a master integral $U$,
\ba
U(x)&\equiv&\int\fr{1}{D_1^xD_2...D_{N}}\;,
\ea
where the $D_i=(p_i^2+m_i^2)$ denote inverse scalar propagators.
In our case the mass parameter $m_i$ has only two values, $0$ and $m$,
the latter of which we
set to $1$, noting that it can be restored in the
end as a trivial dimensional pre-factor of each integral.
The original master integral is then just $U=U(1)$.
Depending on the symmetry properties of the integral,
there can be different choices for the `special' line with
the arbitrary power $x$, but in the limit $x=1$ they all reduce to
the original master integral $U$. This degeneracy can (and will later) 
be used for non-trivial checks of the method.

Employing IBP identities in a systematic way, it is possible
to derive a linear difference equation obeyed by the generalized
master integral $U(x)$,
\ba \label{diffeqn}
\sum_{j=0}^{R} p_{j}(x)U(x+j)&=&F(x)\;, \label{pushdown}
\ea
where $R$ is a finite
positive integer and the coefficients $p_{j}$ are polynomials in $x$
(and the space-time dimension $d$). The function $F$ on the
r.h.s. is a linear combination of
functions analogous to $U(x)$ but derived from `simpler' master integrals,
i.e. integrals containing a smaller number of loops and/or propagators.

The general solution of this kind of an equation is the sum of a special
solution to the full equation, $U_0(x)$, and the solutions to the 
homogeneous equation ($F=0$),
\ba \label{gensoln}
U(x)&=&U_0(x)+\sum_{j=1}^R U_j(x),
\ea
where each ($j=0,...,R$)
\ba \label{soln}
U_j(x)=\mu_j^x \sum_{s=0}^\infty a_j(s) \fr{\Gamma(x+1)}{\Gamma(x+1+s-K_j)}
\ea
is a factorial series\footnote{For a rigorous definition of the concept as
well as a motivation for this kind of an Ansatz, we refer the reader to 
\re\cite{Laporta:2001dd}.}.
Substituting this form into \eq\nr{diffeqn},
one obtains the coefficients $\mu_j$ and $K_j$
(the latter being a function of $d$), as well as recursion relations
for the $x$-independent coefficients $a_j(s)$ (being functions of $d$
as well) for each solution.
For the homogeneous solutions, these recursion relations relate all
coefficients with $s>0$ to their (in principle arbitrary) value at 
$s=0$, $a_j(s)=c_j(s)\,a_j(0)$,
where the $c_j(s)$ are rational functions of $d$.
For the special solution, all $a_0(s)$ are on the other hand completely 
fixed in terms of the inhomogeneous part $F(x)$,
consisting of `simpler' integrals which are assumed to be
already known in terms of their factorial series expansions.

What clearly remains to be done is to fix the $x$\/- and
$s$\/-independent constants $a_j(0)$, $j\neq 0$, in order to
determine the weights of the different homogeneous solutions. To
this end, it is most useful to study the behavior of $U(x)$ at
large $x$. Writing the integral in the form
\ba
U(x)&=&\int\fr1{(p_1^2+1)^x}\, g(p_1),
\ea
it is easy to see that its large-$x$ behavior is determined by the
small-momentum expansion of the two-point function $g(p_1)$, which
has one loop less than the original vacuum integral.

In the case of integrals for which the limit $g(0)$ is well-defined
and non-zero, the calculation becomes particularly simple. Then
the large-$x$ limit of $U(x)$ factorizes into a one-loop
bubble carrying the large power $x$
and a lower-loop vacuum bubble $g(0)$, which corresponds to $U(x)$
with its `special' line cut away,
\ba \label{eq:largeXbc}
\lim_{x\rightarrow\infty}
U(x) &=& \lk \int\fr1{(p_1^2+1)^x}\rk \times \lk \vphantom{\int}
g(0) \rk \;\sim\; (1)^x x^{-d/2} g(0) \;.
\ea
A comparison with
the large-$x$ behavior of \eqs\nr{gensoln}, \nr{soln},
proportional to $\sum_j \mu_j^x a_j(0) x^{K_j}$, can now be used
to fix the $a_j(0)$, of which maximally one will turn out to be
non-zero for our set of integrals.

If on the other hand $g(0)=0$, the treatment of the small-$p_1$
limit of this function becomes more involved. Fortunately, the massless lines
of the sub-diagram --- which were responsible for
the vanishing of its value at zero external momentum in the first place ---
also make its analytic evaluation more straightforward.
Performing a careful analysis
of the subgraph, one always ends up with an integral of the type
\ba
\lim_{x\rightarrow\infty}
U(x) &\sim& \int\fr{(p_1^2)^{\alpha}}{(p_1^2+1)^x},
\ea
from which the calculation proceeds just as above providing us with the 
values of the $a_j(0)$.

Having the full solution at hand, we have in principle completed our
task, as in the limit $x=1$ we recover from $U(x)$ the value
of the initial integral.
Let us, however, add a couple of practical remarks here.
What is still to be done is to perform
the summation of the factorial series of \eq\nr{soln},
which in practice means truncating the infinite sum at some large but 
finite $s_{\rm max}$.
Studying the convergence behavior of these sums, one
notices that even in the cases where they do converge down to $x\sim 1$,
their convergence properties usually strongly decline
with decreasing $x$. This means that in practical computations, where one
aims at obtaining a maximal number of correct digits for
$U(1)$ with as little CPU time as possible, the optimal strategy is to
evaluate the integrals $U(x_{\rm max}+1)$, ..., $U(x_{\rm max}+R)$ with 
the factorial series approach at some $x_{\rm max}\gg 1$ and then use 
the recurrence relation of
\eq\nr{diffeqn} to obtain the desired result at $x=1$.
The price to pay is, however, a loss of numerical
accuracy at each `pushdown' ($x\rightarrow x-1$) step
due to possible cancellations, which
makes the use of a very high $x_{\rm max}$ impossible.
In practice the strategy is to determine an optimal value for the
ratio $s_{\rm max}/x_{\rm max}$.
To give an example, for the four-loop integrals of Section~\ref{se:four}
we have found that $s_{\rm max}/x_{\rm max}\sim 20...40$ is a good value,
while we used a range of $s_{\rm max}\sim 2000\dots 2500$. 
For a few special cases,
for which additional numerical problems emerged, we were forced to limit the
value of the parameters to roughly $s_{\rm max}\sim 200$ and 
$x_{\rm max}\sim 30$, 
which decreased the accuracy of the results significantly.

%%%%%%%%%%%%%%%%%%%%%%%%%%%%%%%%%%%%%%%%%%%%%%%%%%%%%%%%%%%%
%%%%%%%%%%%%%%%%%%%%%%%%%%%%%%%%%%%%%%%%%%%%%%%%%%%%%%%%%%%%
%%%%%%%%%%%%%%%%%%%%%%%%%%%%%%%%%%%%%%%%%%%%%%%%%%%%%%%%%%%%

\section{\label{se:three} Implementation of the algorithm}

As is apparent from the preceding section, there are three main steps
involved in obtaining the desired numerical coefficients in the
$\e$\/-expansion of each master integral:
deriving the difference equations obeyed by each integral,
solving them in terms of factorial series,
and finally performing the $\e$\/-expansion and numerically
evaluating the sum of \eq\nr{soln} (truncated at $s_{\rm max}$)
to the precision needed. We will briefly address each of them in the 
following.

For the first step, we slightly generalized the IBP algorithm
we had used for reducing generic 4-loop bubble integrals to master
integrals, which follows the setup given in \re\cite{Laporta:2001dd},
and whose implementation in FORM \cite{Vermaseren:2000nd}
is documented in \re\cite{Schroder:2002re}.
The main difference is an enlarged representation for the
integrals, keeping track of the line which carries the
extra powers $x$, as well as the fact that there are now
two independent variables ($d$, $x$), requiring factorization
(and inversion) of bivariate polynomials, as opposed to univariate
polynomials in the original version.

Second, staying within FORM for convenience, we implemented routines
that straightforwardly solve the difference equations in terms
of factorial series, along the lines of \re\cite{Laporta:2001dd}.
This is done starting with the simplest one-loop master integral,
and working the way up to the most complicated (most lines)
four-loop integral,
ensuring that at each step, the `simpler' terms constituting the
inhomogeneous parts of the difference equation are already known.
The output are then plain ascii files
specifying each solution in the form of \eq\nr{soln} as well as
containing recursion relations for the coefficients $a(s)$.
Note that these first two steps are performed exactly, in $d$
dimensions.

Third, once the recursion relations for the coefficients $a(s)$ were known,
we used a Mathematica program to obtain their numerical values at each
$s$ to a predefined precision, and to perform the summation of the
factorial series.
While this procedure is in principle straightforward,
there are some twists that we employed to help reduce the
running times significantly, most of
which are probably
quite specific to our use of Mathematica.
To avoid a rapid loss of significant digits in
solving the recursion steps that relate each $a(s)$ to $a(0)$, especially
those for the homogeneous coefficients, we first solved the relations
analytically and only in the end substituted the
numerical value (actually the truncated $\e$\/-expansion)
of the first non-zero coefficient.
In fact, we found Mathematica to operate quite efficiently
with operations like multiplication of two truncated power series,
so that we relied heavily on it.
Furthermore, since --- not surprisingly --- the
most time-consuming part in the summation of the series
turned out to be the $\e$\/-expansion of $\Gamma$\/-functions,
we achieved a notable speed-up by substituting the $\Gamma$\/-functions with
large arguments by suitable products of linear factors times
$\Gamma$\/-functions of smaller arguments.
Finally, a vital step in avoiding an excessive loss in the
depth of the $\e$\/-expansions when going
from one integral to the next, was to apply the
`Chop' command to remove from the results and coefficients excess
unphysical poles, whose coefficients were of the
order of, say, $10^{-50}$ or less. In some cases we were in addition
able to reduce the loss of precision in the pushdown steps by first 
analytically solving $U(1)$ as a function of $U(x_{\rm max})$, and only 
in the very end substituting the numerical value of the latter.

%%%%%%%%%%%%%%%%%%%%%%%%%%%%%%%%%%%%%%%%%%%%%%%%%%%%%%%%%%%%
%%%%%%%%%%%%%%%%%%%%%%%%%%%%%%%%%%%%%%%%%%%%%%%%%%%%%%%%%%%%
%%%%%%%%%%%%%%%%%%%%%%%%%%%%%%%%%%%%%%%%%%%%%%%%%%%%%%%%%%%%

\section{\label{se:four} Numerical results}

Below we list the Laurent expansions in $\e=(4-d)/2$ of 
vacuum master integrals up to four loops. We use an intuitive
graphical notation, in which each solid line represents a massive
scalar propagator $1/(p^2+1)$ and a dashed line a massless one
$1/p^2$. 
The integral measure we have chosen here is
\ba
\int_p &\equiv& \fr{1}{\Gamma(2+\e)}\int\!
\fr{{\rm d}^{4-2\e}p}{\pi^{2-\e}}\;,
\ea
which implies that the 1-loop tadpole is 
$J=\int_p\frac1{p^2+1}=\frac{-1}{\e(1-\e^2)}=-\sum_{n=0}^\infty \e^{2n-1}$.
In each case we provide the results to order $\e^{10}$ keeping
the accuracy at 50 significant digits for the 2- and 3-loop
master integrals and at 40 for the 4-loop ones. 
There are two exceptions. For one of the 3-loop integrals
(see \eq\nr{Mh} below) the factorial series does not
converge and hence the integral has to be treated by Laplace transform,
see Section~\ref{se:Laplace}. 
For one of the 4-loop integrals (see \eq\nr{Xa} below)
we only give the first seven $\e$\/-orders to 17 significant digits.
To obtain more
$\e$\/-orders and significant digits for all integrals listed here 
is merely a matter of additional CPU time.

We have produced numerical results for all single-mass-scale
vacuum master integrals up to three loops
(these are the master integrals entering the package of Avdeev 
\cite{Avdeev:1995eu} and MATAD \cite{Steinhauser:2000ry}), 
and for all `QED-type' vacuum master integrals at four loops.
Here, we display only those numerical results which correspond neither 
to analytically solvable integrals (1 of 1 1-loop master, 
1 of 2 2-loop masters, 3 of 12 3-loop masters, 2 of 10 4-loop masters,
all of which are given analytically in Section~\ref{se:Analytic} below, 
and are listed in numerical form in the appendix), 
nor to fully massive cases (1 of 2 2-loop masters, 
3 of 12 3-loop masters, which are given in 
Section~9.3.1 of \re\cite{Laporta:2001dd};
1 of 10 4-loop masters, 
which is given in \eq(4) of \re\cite{Laporta:2002pg}).

\ba
\threeBb &=&{} \label{Bb}
+1.0000000000000000000000000000000000000000000000000\ep^{-3}
\nn&&{}+0.75000000000000000000000000000000000000000000000000\ep^{-2}
\nn&&{}+2.8750000000000000000000000000000000000000000000000\ep^{-1}
\nn&&{}+1.8362912870512825038535151499626234054646567716807
\nn&&{}-26.427828097688527267319254765120590367456377175480\ep
\nn&&{}-35.088051385481306364065961402117419432373775682177\ep^2
\nn&&{}-512.75537623727044689027104289971864365971796649684\ep^3
\nn&&{}-607.61494953927726782115473332930225595551912885034\ep^4
\nn&&{}-5868.5987295458313170081280447224279031237930577453\ep^5
\nn&&{}-6835.6108788455114123641492279253803965001408075543\ep^6
\nn&&{}-58194.090725773231428299235587057139067942816045554\ep^7
\nn&&{}-67435.335245041201792055506063164867635607825896649\ep^8
\nn&&{}-546094.78026628005592146280450252032502449454982782\ep^9
\nn&&{}-631563.41278231233491152773645513360004834876043263\ep^{10}
       +\order{\e^{11}}\\
\threeVb&=&{} \label{Vb}
-0.66666666666666666666666666666666666666666666666667\ep^{-3}
\nN&&{}-1.6666666666666666666666666666666666666666666666667\ep^{-2}
\nn&&{}-5.1290732088140381005700717333376095855758453867530\ep^{-1}
\nn&&{}-26.359970069205366659319388577532454678949629074714
\nn&&{}-27.711175418518951962132692178901111387631205211816\ep
\nn&&{}-293.15661097603756640443665077615751632698451158842\ep^2
\nn&&{}-142.70296384808301760189570443963964069968530061393\ep^{3}
\nn&&{}-2882.1838924952422595902727649575335612132315437366\ep^{4}
\nn&&{}-801.64629651874722343778241421866459175486305997074\ep^{5}
\nn&&{}-26947.975116190322227046885628024191588708203470044\ep^{6}
\nn&&{}-5202.1954102253813787831194867097139379256134207497\ep^{7}
\nn&&{}-246612.58893683893330836716807041349553641918919392\ep^{8}
\nn&&{}-38662.312198716830334636275721442722552625311137805\ep^{9}
\nn&&{}-2235893.9169450155346378997842831790622571843918826\ep^{10}
       +\order{\e^{11}}\\
\threeMc &=&{} \label{Mc}
+2.4041138063191885707994763230228999815299725846810\ep^{-1}
\nN&&{}-13.125546202841586242894146861604104971473328745577
\nn&&{}+58.026260003878655576719786597271170572487856789112\ep
\nn&&{}-215.15799420773251496754795359758001229751012774168\ep^{2}
\nn&&{}+741.02167568175382570477503744319405056068752490840\ep^{3}
\nn&&{}-2422.8745603243623464433277838674972111388328822910\ep^{4}
\nn&&{}+7691.0946660371679072096695375419004253731139722117\ep^{5}
\nn&&{}-23935.477541938694107878632636617038283240279946231\ep^{6}
\nn&&{}+73567.948130076321368433008921329180208448045889323\ep^{7}
\nn&&{}-224259.22429731742234354745849077250951082319381698\ep^{8}
\nn&&{}+679949.06185664528482517935972319009053625673803897\ep^{9}
\nn&&{}-2054250.6137900838709156880585181458050495843111626\ep^{10}
       +\order{\e^{11}}\\
\threeMf &=&{} \label{Mf}
+2.4041138063191885707994763230228999815299725846810\ep^{-1}
\nN&&{}-10.073203643096893062671213536841941862151359216063
\nn&&{}+46.082030897278984204342981632818973100797752268016\ep
\nn&&{}-162.84321571472549604685427998929495247564452607777\ep^{2}
\nn&&{}+563.02541599052549921690912303391142482056503193963\ep^{3}
\nn&&{}-1822.8278416039379661792322993085062379900421439244\ep^{4}
\nn&&{}+5785.9815122286472118701238800303861843636492370028\ep^{5}
\nn&&{}-17968.847688521304415691142884872355494614421789034\ep^{6}
\nn&&{}+55216.376506037509642111329809657803343085975716148\ep^{7}
\nn&&{}-168240.56307714987438328576061703042187348961355271\ep^{8}
\nn&&{}+510052.27830760883492002666904963035268124951366379\ep^{9}
\nn&&{}-1540802.7858406456522499592944279183737583148997338\ep^{10}
       +\order{\e^{11}}\\
\threeMh &=&{} \label{Mh} +2.40411380631919\ep^{-1}
       -6.09209302191832
\nN&&{}+35.8130598712514\ep
       -104.744695525740\ep^{2}
\nn&&{}+394.7643404810\ep^{3}
       -1200.978166746\ep^{4} +\order{\e^{5}}\\
\threeMe &=&{} \label{Me}
+2.4041138063191885707994763230228999815299725846810\ep^{-1}
\nN&&{}-10.239350912945217732184803670827657230740659540460
\nn&&{}+46.310233388509835938575195677581891346572247104081\ep
\nn&&{}-163.71903666846274587940160817767510251267798801563\ep^{2}
\nn&&{}+564.30910069499791449917891053192483169414448523830\ep^{3}
\nn&&{}-1825.8206691490586101339592917414250683553208417056\ep^{4}
\nn&&{}+5790.4830503256226500389801844087331481909799175043\ep^{5}
\nn&&{}-17977.329926014373828927297954140399343188401436964\ep^{6}
\nn&&{}+55229.228709840982549894271961274899889724333654202\ep^{7}
\nn&&{}-168262.33984881039469900336415583383277506556682677\ep^{8}
\nn&&{}+510085.27212468040781599787353617454516173782111015\ep^{9}
\nn&&{}-1540855.6379207615212777547545938657889645397412471\ep^{10}
       +\order{\e^{11}}\\
\fourBBb &=&{} \label{BBb}
-1.000000000000000000000000000000000000000\ep^{-4}
\nN&&{}-0.500000000000000000000000000000000000000 \ep^{-3}
\nn&&{}-3.527777777777777777777777777777777777778 \ep^{-2}
\nn&&{}-1.995370370370370370370370370370370370370 \ep^{-1}
\nn&&{}-36.82021604938271604938271604938271604938
\nn&&{}-19.87920801451107035069575635156380101575 \ep
\nn&&{}-1809.001126638637160933894507798781706682 \ep^2
\nn&&{}-941.2486498215135407529753614624254521594 \ep^3
\nn&&{}-49114.80404240275263940837370626747663512 \ep^4
\nn&&{}-25712.87944658606239888301931387195377680 \ep^5
\nn&&{}-1014742.540337323108931396794699794706304 \ep^6
\nn&&{}-533925.9315185165221824312193117157135164 \ep^7
\nn&&{}-18513953.44519328478360685998151320048728 \ep^8
\nn&&{}-9769845.146715270007449428486953016122496 \ep^9
\nn&&{}-317669932.9515691976277658362596784695115 \ep^{10}
       +\order{\e^{11}}\\
\fourTb &=&{} \label{Tb}
+0.2500000000000000000000000000000000000000\ep^{-4}
\nN&&{}+0.5000000000000000000000000000000000000000\ep^{-3}
\nn&&{}+1.000000000000000000000000000000000000000\ep^{-2}
\nn&&{}+1.813369870537362855098298049824424939972\ep^{-1}
\nn&&{}-113.8224542836131461311762552843948945680
\nn&&{}-33.70692008121875082746730709549318292582\ep
\nn&&{}-3800.131177952398833364468086486701310324\ep^{2}
\nn&&{}-724.2483980459868435529785916706580415218\ep^{3}
\nn&&{}-83243.75114211351557600351242603548310943\ep^{4}
\nn&&{}-9962.244874731471054690629554209449745080\ep^{5}
\nn&&{}-1556494.392681571176934758495668112116219\ep^{6}
\nn&&{}-125852.9269007094630774780949002157883896\ep^{7}
\nn&&{}-27026768.74324139691004925806420463865625\ep^{8}
\nn&&{}-1619900.985945231199760429618558131115494\ep^{9}
\nn&&{}-451968203.1707264233126870326577507342793\ep^{10}
       +\order{\e^{11}}\\
\fourTc &=&{} \label{Tc}
+0.6666666666666666666666666666666666666667\ep^{-4}
\nN&&{}+1.333333333333333333333333333333333333333\ep^{-3}
\nn&&{}+3.333333333333333333333333333333333333333\ep^{-2}
\nn&&{}-2.922363183148830477868063138605600049253\ep^{-1}
\nn&&{}-52.50529739842769756973487794955803028226
\nn&&{}-622.1548972708590376012515685880304077291\ep
\nn&&{}-1741.392944346262052260405956917114927201\ep^{2}
\nn&&{}-17196.12685330902582768340098554237636824\ep^{3}
\nn&&{}-35037.76438140725371856293904191777384497\ep^{4}
\nn&&{}-350040.6052285494016783912074340410365119\ep^{5}
\nn&&{}-619669.7756160060500505884016704452111642\ep^{6}
\nn&&{}-6316632.078794015469602341973684043315269\ep^{7}
\nn&&{}-10420684.66626276045383010214928284087492\ep^{8}
\nn&&{}-107682720.9936656086807201002498313447872\ep^{9}
\nn&&{}-171175743.7334785889316026497774359587050\ep^{10}
       +\order{\e^{11}}\\
\fourVBb &=&{} \label{VBb}
-0.1666666666666666666666666666666666666667\ep^{-4}
\nN&&{}-0.8333333333333333333333333333333333333333\ep^{-3}
\nn&&{}-5.535390236492927618733071494844783324098\ep^{-2}
\nn&&{}-18.82211358179364443034084677047078519365\ep^{-1}
\nn&&{}-25.33131709639103630297934297632219102642
\nn&&{}-692.6253681383859207802291611352811358818\ep
\nn&&{}+1304.406827189023173835521731683389467596\ep^{2}
\nn&&{}-17597.62761742767175796342110253040416842\ep^{3}
\nn&&{}+43608.68478725040973761321535022863250602\ep^{4}
\nn&&{}-356925.7947952212585233385307804100264907\ep^{5}
\nn&&{}+939175.5208936133499000732308171881559393\ep^{6}
\nn&&{}-6467516.567931160982387324881460909595434\ep^{7}
\nn&&{}+17364082.00316543469946942544036134544207\ep^{8}
\nn&&{}-110630064.0504718962799294108969364410321\ep^{9}
\nn&&{}+299555848.2967199841801845664112146544429\ep^{10}
       +\order{\e^{11}}\\
\fourVBa &=&{} \label{VBa}
-0.1666666666666666666666666666666666666667\ep^{-4}
\nN&&{}-0.8333333333333333333333333333333333333333\ep^{-3}
\nn&&{}-4.934361784913130476033202414089058328716\ep^{-2}
\nn&&{}-16.20728580177089935860423675579773428042\ep^{-1}
\nn&&{}-66.26045267088253719950145322703822192330
\nn&&{}-375.8131807648258568590885739785987842170\ep
\nn&&{}-558.8684980291056611497739327005648555056\ep^{2}
\nn&&{}-8005.909258052308131324607228836046381631\ep^{3}
\nn&&{}-1919.455472313357401527573620257269458394\ep^{4}
\nn&&{}-151678.4473872174906037102434226085312395\ep^{5}
\nn&&{}+45367.13703676553118943884640644928312337\ep^{6}
\nn&&{}-2666691.771475554201053771443874777744895\ep^{7}
\nn&&{}+1469487.259313641787542428480707831779115\ep^{8}
\nn&&{}-44948193.38776277686896804573212402472750\ep^{9}
\nn&&{}+30458142.54983328875970186426762508169587\ep^{10}
       +\order{\e^{11}}\\
\fourWa &=&{} \label{Wa}
+5.184638775716849631656827432285170840285\ep^{-1}
\nN&&{}-43.27615856932464061120186605318440014787
\nn&&{}+281.6878028207571441503294898731325032572\ep
\nn&&{}-1513.439498357334510783856768928671867962\ep^{2}
\nn&&{}+7425.188218180801392788420314778806091908\ep^{3}
\nn&&{}-34157.28328369996440389044714608216014037\ep^{4}
\nn&&{}+150927.1863992836861076852919478736657397\ep^{5}
\nn&&{}-648212.2200729992344766408701706025837197\ep^{6}
\nn&&{}+2730180.324952706549742379412290749026850\ep^{7}
\nn&&{}-11339683.06464037751511038174569497508966\ep^{8}
\nn&&{}+46630964.11789747669532801502683952077042\ep^{9}
\nn&&{}-190369535.4103429881202484621694879170262\ep^{10}
       +\order{\e^{11}}\\
\fourXa &=&{} \label{Xa}
+1.80887954620833474
       -12.7814836099524403\ep
\nN&&{}+71.046049240835262\ep^{2}
       -334.59648933739741\ep^{3}
\nn&&{}+1467.5837602507405\ep^{4}
       -6165.5621168597119\ep^{5}
\nn&&{}+25329.619267580422\ep^{6}
       +\order{\e^{7}}
\ea

Just as in the three-dimensional case \cite{Schroder:2003kb}, 
we have performed
various checks on our results. These can be divided into two
categories: first, we had to make sure that the difference equation,
\eq\nr{diffeqn}, as well as the various parts of its solution,
\eqs\nr{gensoln}, \nr{soln}, were
in principle correct, and second, that we have
reached the desired accuracy in the numerical part of our computation. 
For the first task, it was in general enough to ensure that the first 
few $\e$-orders we obtained for each
integral coincided with the existing analytic results. Here the main
difference to our previous three-dimensional computation laid in the
fact that while at $d=3-2\e$ most of the integrals were either finite
or their expansions started with a $1/\e$ term, we now encountered in many
cases (often analytically calculable) divergent terms up to $1/\e^4$
order.
The analytic results relevant to our graphs that we have
found in the literature, as well as a few new ones, 
are collected in Section~\ref{se:Analytic}.
We have found agreement in all cases.

The comparisons with existing analytic results also provide an easy 
and reliable method to inspect the
accuracy of the numerical results, since the number of correct digits 
usually stays roughly constant when
moving from one $\e$-order to the next. Just as in our previous work 
\cite{Schroder:2003kb}, other methods we have employed to assess the 
accuracy question include
comparing the results obtained by raising topologically inequivalent 
lines in a single integral to a higher power and analyzing the convergence
properties of the factorial series, i.e.~checking the stability of our 
results with respect to varying $s_{\rm max}$. The
results given in the preceding section have been observed to be stable 
to at least the number of digits shown.

One might be concerned about the rapid growth with increasing
$\e$\/-orders of most of the coefficients.
This is, as was pointed out in \re\cite{Laporta:2002pg}, caused
by poles that the integrals (seen as functions of $d$) develop near
$d=4$, e.g. at $d=7/2, 3$, etc. It is to be expected that factoring out 
the first few of these nearby
poles in each case will improve the apparent convergence in $\e$
considerably.

In principle, having a method at hand that is capable of
generating coefficients to very high accuracy, even to a couple
of hundred digits, one could now use the algorithm PSLQ \cite{pslq}
combined with an educated guess of the number content of some
of the yet-unknown constant terms, in order to search for
analytic representations of the numerical results. 
We have not made any systematic attempts in that direction, since
the numerical accuracy of our results should be sufficient for
all practical purposes.

%%%%%%%%%%%%%%%%%%%%%%%%%%%%%%%%%%%%%%%%%%%%%%%%%%%%%%%%%%%%
%%%%%%%%%%%%%%%%%%%%%%%%%%%%%%%%%%%%%%%%%%%%%%%%%%%%%%%%%%%%
%%%%%%%%%%%%%%%%%%%%%%%%%%%%%%%%%%%%%%%%%%%%%%%%%%%%%%%%%%%%

\section{\label{se:Laplace} Laplace transform}

As already mentioned in the above, we have encountered one case
where the method of computing the $\e$\/-expansion
via a factorial series representation does not work
(or, more precisely, does not converge),
namely for the 3-loop integral of \eq\nr{Mh}.
Let us take this specific example as an opportunity 
to finally display a difference equation like \eq\nr{diffeqn}
in full detail, and exhibit, following \re\cite{Laporta:2001dd}, 
one method other than factorial series for solving it.

Defining the integral
\ba
\Mh(x) 
&\equiv& \frac{\threeMhx}{\twoSb\oneJ} 
\;=\; \frac{\threeMhx}{J^3} \frac{2^{d-2}\Gamma(\frac12)}
{\Gamma(\frac{3-d}2)\Gamma(\frac{d}2)}\;,
\ea
where the dot with the label $x$ means that the corresponding
propagator is raised to the $x$\/-th power,
the difference equation \eq\nr{diffeqn} it satisfies 
is of second order and reads
\ba
\label{eq:deqMh}
0 &=& -2(x+1)\Mh(x+2)+3(x+2-d/2)\Mh(x+1)-(x+3-d)\Mh(x) \nn&&{}
+\frac{\Gamma(x+5-\frac{3d}2)}{\Gamma(x+1)} 
\frac{3-d}{\Gamma(5-\frac{3d}2)} \Mh(0)  
+\frac{\Gamma(x+3-d)}{\Gamma(x+1)} \frac1{\Gamma(2-d)} \nn&&{}
-\frac{\Gamma(x+2-\frac{d}2)}{\Gamma(x+1)} \frac2{\Gamma(1-\frac{d}2)}
+\frac{\Gamma(x+5-\frac{3d}2)\Gamma(x+3-d)}{\Gamma(x)\Gamma(x+7-2d)} 
\frac2{\Gamma(1-\frac{d}2)} \;,
\ea
with boundary conditions 
$\Mh(x\gg1)\sim x^{-\frac{d}2}$ (cf. eq\nr{eq:largeXbc}) and
\ba
\Mh(0) &=& \frac{\threeVe}{\twoSb\oneJ} 
\;=\; -\frac{\Gamma(\frac{3d}2)\Gamma(1-\frac{3d}2)\Gamma(\frac{d}2-1)
\Gamma(\frac{d}2)}{\Gamma(d)\Gamma(d-2)\Gamma(1-d)} \;.
\ea

We would like to know the master integral $\Mh(1)$,
or at least its $\e$\/-expansion in $d=4-2\e$ dimensions. 
Note that only the first two terms of that expansion
are known, cf. \eq\nr{eq:MhAnal} below.
Formally, it is of course possible to solve \eq\nr{eq:deqMh}
in terms of factorial series, following the recipe sketched in
Section~\ref{se:two}. 
However, it turns out that the series does not converge in this
(and only this, of all cases treated in this paper) case, 
such that in practice a different
method of solving the difference equation is needed.

One way to tackle \eq\nr{eq:deqMh} could be the iterative method
used e.g. in \re\cite{Moch:2004sf} (cf. \eq(11)ff therein): 
expand in $\e$, make Ans\"atze for the 
$\e$-coefficients of $\Mh$ in terms of (sums of multiple) harmonic sums
with unknown constants,
rewrite all $\e$-expansions of the Gamma functions in terms of
harmonic sums, then rewrite everything in terms of a unique basis,
and finally fix the constants by comparing coefficients.
Unfortunately, there does not seem to exist an algorithm yet that automatizes
the choice of Ansatz, hence requiring a fair amount of hand-work. 
For the basic literature on harmonic sums, see
the references of \re\cite{Moch:2004sf}.

Another way of tackling \eq\nr{eq:deqMh} is to transform it to 
a differential equation, which should then be solved by analytical
or numerical methods, or by a combination of both. This is what
we will do in the following, and this is how we have obtained the 
numerical values given in \eq\nr{Mh}.

Following \re\cite{Laporta:2001dd}, we can Laplace transform the 
difference equation for $\Mh$,
making the Ansatz $\Mh(x)=\int_0^1 dt\, t^{x-1} v(t)$,
into a first order differential equation 
$\Phi_0(t)v(t)-t\Phi_1(t)v'(t)=w(t)$,
where $\Phi_0(t)=3-d-3(1-\frac{d}2)t-2t^2$ and 
$\Phi_1(t)=(1-t)(1-2t)$.

The homogeneous equation is solved by 
$v_H(t)=c_H t^{3-d}(1-t)^{\frac{d}2-2}(1-2t)^{\frac{d}2-2}$,
which however makes $\Mh(x\gg1)$ grow too fast at large $x$
(it would grow like $x^{1-\frac{d}2}$, in conflict with the large-$x$ 
boundary condition),
such that $c_H\equiv0$ and hence $\Mh^H(x)=0$.

For solving the inhomogeneous equation, note that the inhomogeneous piece 
has four terms
$w(z)=\sum_{j=1}^4 w_j(z)$, which correspond to the last four terms of
\eq\nr{eq:deqMh}, written as $T_j(x)=\int_0^1 dz\,z^{x-1} w_j(z)$.
For $j=1,2,3$ we therefore have 
$w_j(z)=\frac{b_j}{\Gamma(1-a_j)} z^{a_j}(1-z)^{-a_j}$,
where $\vec a=(5-\frac{3d}2,3-d,2-\frac{d}2)$ and 
$\vec b=(\frac{(3-d)\Mh(0)}{\Gamma(5-\frac{3d}2)},\frac1{\Gamma(2-d)},
-\frac2{\Gamma(1-\frac{d}2)})$.
For $w_4$, we know that it satisfies  
\ba \label{eq:bcw4}
\int_0^1 dz\, z^{x-1} w_4(z) &=&
T_4(x) \;=\; 
\frac{\Gamma(x+5-\frac{3d}2)\Gamma(x+3-d)}{\Gamma(x)\Gamma(x+7-2d)} 
\frac2{\Gamma(1-\frac{d}2)} \;.
\ea
For $T_4(x)$, there obviously is a simple difference equation,
$x(x+7-2d)T_4(x+1)=(x+5-\frac{3d}2)(x+3-d)T_4(x)$, from which we 
get -- in complete analogy to Laplace transforming \eq\nr{eq:deqMh} --
a differential equation for $w_4$:
\ba
0 &=& 
(d-3)(3d-10-4z)w_4(z)-z(14-5d+4z(d-2))w_4'(z)+2z^2(1-z)w_4''(z) \;.
\ea
To write its boundary condition \eq\nr{eq:bcw4} 
in a (for numerical treatment) more useful form,
note that the behavior of $w_4(z)$ at the singular point $z=1$
is connected to the large-$x$ limit of $T_4(x)$. Using Stirling's
formula to write $\frac{\Gamma(x+a)}{\Gamma(x+b)}=x^{a-b}
(1+\frac{(a-b)(a+b-1)}{2x}+\order{x^{-2}})$, we can fix the three
constants in the Ansatz 
$w_4(z\approx1)=c_1 (1-z)^{c_2}(1+c_3(1-z)+\dots)$,
when comparing $\int_0^1 dz\,z^{x-1}w_4(z\approx1)$ at large $x$
with $T_4(x\gg1)$.
Hence, writing 
\ba \label{eq:w4barDef}
w_4(z) \;=\; 
c_1 (1-z)^{c_2} \frac{2 z^{1-\frac{d}2}}{d-2} \bar{w}_4(z)\;=\; 
\frac{4(1-z)^{\frac{d}2-2}z^{1-\frac{d}2}\bar{w}_4(z)}
{\Gamma(1-\frac{d}2)\Gamma(\frac{d}2-1)(d-2)}\;=\; 
\frac{2\sin\frac{\pi d}2}{\pi}\,z^{1-\frac{d}2} 
(1-z)^{\frac{d}2-2} \bar{w}_4(z)\;,
\ea
we get simple boundary conditions $\bar{w}_4(1)=\frac{d-2}2$, 
$\bar{w}_4'(1)=\frac{d-2}2(\frac{d-2}2-c_3)
=\frac{(d-4)^2}2$ for the new function $\bar{w}_4(z)$, which
satisfies the differential equation
\ba \label{eq:w4bar}
0 &=& -(d-4)^2\bar{w}_4(z)
+z(10-3d+4z(d-3))\bar{w}_4'(z)+2z^2(z-1)\bar{w}_4''(z) \;.
\ea

We now get the non-homogeneous solution $v_{NH}(t)$ by varying the constant 
of the homogeneous solution. 
Due to the linearity of the differential equation, the full solution
is simply the sum of four terms, which when plugged back into the
definition of the Laplace transform gives a representation
for the master $\Mh$:
\ba \label{eq:Mhsoln}
\Mh(x) &=& \int_0^1 dt\, t^{x+2-d} (1-t)^{\frac{d}2-2} (1-2t)^{\frac{d}2-2}
\int_t^1 dz\, z^{d-4} (1-z)^{1-\frac{d}2} (1-2z)^{1-\frac{d}2}\times \nn&&{}
\times\lb \sum_{j=1}^3 \frac{b_j}{\Gamma(1-a_j)} z^{a_j}(1-z)^{-a_j} 
+w_4(z)\rb \;.
\ea
The integral converges (in 4d) for $x>1$, so one can use it to
compute $\Mh(2)$ and get $\Mh(1)$ via \eq\nr{eq:deqMh}.

Unable to solve \eq\nr{eq:Mhsoln} for generic $d$, 
let us now go to $d=4-2\e$ dimensions and start expanding.
First, we need to solve the differential equation
\eq\nr{eq:w4bar}.
Writing $\bar{w}_4(z)=\sum_{n=0}^\infty \e^n\, f_n(z)$, 
the boundary conditions translate into 
$f_0(1)=1$, $f_1(1)=-1$, $f_{n>1}(1)=0$ and
$f_0'(1)=0$, $f_1'(1)=0$, $f_2'(1)=2$, $f_{n>2}'(1)=0$.
The $f_n(z)$ satisfy the differential equations
\ba \label{eq:fn}
0 &=& z(z-1)f_n''(z) +(2z-1)f_n'(z) 
+(3-4z)f_{n-1}'(z)-\frac2z f_{n-2}(z) \;,
\ea
which have to be solved starting with $n=0$ (setting $f_{n<0}(z)\equiv0$).

One can e.g. solve \eq\nr{eq:fn} in terms of multiple integrals.
The Ansatz $f_n(z)=\delta_{n,0}-\delta_{n,1}+\int_1^z da\, g_n(a)$
respects the boundary conditions for $f_n(1)$ and
transforms \eq\nr{eq:fn} into a first order differential equation 
for $g_n(a)$, whose boundary conditions 
$g_n(1)=2\delta_{n,2}$ incorporates those for $f_n'(1)$. 
The homogeneous solution is of the form 
$g_n^H(a)=\frac{c_n}{a(1-a)}$ and vanishes
due to the boundary condition: $c_n\equiv0$.
The inhomogeneous solution now follows by variation of the constant,
such that finally
\ba
\label{eq:fnn}
f_n(z) &=& \delta_{n,0}-\delta_{n,1}+\int_1^z \frac{da}a 
\frac{h_n(a)}{1-a} \;,\\
\label{eq:hn}
h_n(a) &=& 2(\delta_{n,3}-\delta_{n,2})\ln(a)
+\int_1^a \frac{db}b \frac{(3-4b)h_{n-1}(b)}{1-b}
-2 \int_1^a \frac{db}b \int_1^b \frac{dc}c \frac{h_{n-2}(c)}{1-c} \;.
\ea

The strategy is now clear: $h_n(a)\rightarrow f_n(z)\rightarrow\bar w_4(z)
\rightarrow w(z) \rightarrow \Mh(2) \rightarrow \Mh(1)$.
All of these steps can be done numerically, and there is a discussion
of practical methods in \re\cite{Laporta:2001dd}.

In practice, we numerically solved for the $f_n$ using
Mathematica, changed the order of integrations in \eq\nr{eq:Mhsoln}, 
dealt with the $t$\/-integration (semi-) analytically, and finally 
performed the $z$\/-integration numerically. 
The singular point at $z=1/2$ was treated as a
pricipal value integral, and the logarithmically divergent regions near
$z=1/2$ and $z=1$ were split off
and treated analytically via a series-expansion in $z$.

To check the setup, it is possible to start analytically.
Solving \eq\nr{eq:hn}, the first couple of orders for $h_n$ read 
$h_0(a)=0$, 
$h_1(a)=0$, 
$h_2(a)=-2\ln(a)$ and
$h_3(a)=2\ln(a)-3\ln^2(a)+2\li_2(1-a)$,
where $\li_n(z)=\sum_{k=1}^\infty \frac{z^k}{k^2}$ is the polylogarithm.

Using \eq\nr{eq:fnn}, this then implies
$f_0(z)=1$,
$f_1(z)=-1$,
$f_2(z)=-\ln^2z-2\li_2(1-z)$,
$f_3(z)=(1+5\ln(1-z)-\ln(z))\ln^2(z)+2(1+\ln(z))\li_2(1-z)
+10\ln(z)\li_2(z)-2\li_3(1-z)-10\li_3(z)+10\zeta_3$.

Knowing now $\bar{w}_4(z)=1-\e+\e^2 f_2(z)+\order{\e^3}$
and using \eq\nr{eq:w4barDef},
we can expand the curly bracket of \eq\nr{eq:Mhsoln}. 
The two leading terms cancel, such that
$\lb..\rb=\frac{2\e^3}{3z}[(z-1)(\pi^2+2\ln^2(1-z)-6\ln
z\ln(1-z))+3z \ln^2z+6\li_2(1-z)]+\order{\e^4}$. 
Now $\Mh(x)=\int_0^1 dt\, t^{x-2} \int_t^1 dz\, \frac{\lb..\rb}
{(1-z)(1-2z)}+\order{\e^4}=\int_0^1 dz\, \frac{\lb..\rb}{(1-z)(1-2z)}
\frac{z^{x-1}}{x-1}+\order{\e^4}$. 
We obtain $\Mh(2)=6\zeta_3 \e^3+\order{\e^4}$,
which, using \eq\nr{eq:deqMh} at $x=0$, translates into 
$\Mh(1)=\frac4{3(4-d)}\Mh(2)=\frac2{3\e}\Mh(2)=4\zeta_3 \e^2+\order{\e^3}$,
in nice agreement with the first term of \eq\nr{eq:MhAnal}.

%%%%%%%%%%%%%%%%%%%%%%%%%%%%%%%%%%%%%%%%%%%%%%%%%%%%%%%%%%%%
%%%%%%%%%%%%%%%%%%%%%%%%%%%%%%%%%%%%%%%%%%%%%%%%%%%%%%%%%%%%
%%%%%%%%%%%%%%%%%%%%%%%%%%%%%%%%%%%%%%%%%%%%%%%%%%%%%%%%%%%%

\section{\label{se:Analytic} Analytic results}

For completeness we list here all existing analytic results
applicable to our integrals that we are aware of.

Here, we normalize every integral with the appropriate
power of the 1-loop tadpole, such that analytic results are independent
of the integration measure.
Also, recall that we have set $m=1$.

We will use the following transcendentals:
\ba
\zeta_n &=& \sum_{k=1}^\infty \frac1{k^n} \;,\\
a_n &=& \sum_{k=1}^\infty \frac1{2^k k^n} \;=\; \li_n(1/2) \;,\\
\ls_j(\theta) &=& \label{eq:ls}
 -\int_0^{\theta} d\tau \ln^{j-1}|2\sin\frac\tau2| \;,
\ea
and abbreviate the log-sine integrals at their maximum value as
$\ls_j(\frac{2\pi}3)\equiv\ls_j$ below.

%%%%%%%%%%%%%%%%%%%%%%%%%%%%%%%%%%%%%%%%%%%%%%%%%%%%%%%%%%%%
%%%%%%%%%%%%%%%%%%%%%%%%%%%%%%%%%%%%%%%%%%%%%%%%%%%%%%%%%%%%

\subsection{1-loop}

There is one 1-loop topology and one coloring by mass.
The 1-loop tadpole has an analytic solution in terms of Gamma functions.
With measure $\int d^dp$, $J=\int d^dp \frac1{p^2+1}=\pi^{d/2}\Gamma(1-d/2)$.
\ba
\oneJ &\equiv& J \;.
\ea

%%%%%%%%%%%%%%%%%%%%%%%%%%%%%%%%%%%%%%%%%%%%%%%%%%%%%%%%%%%%
%%%%%%%%%%%%%%%%%%%%%%%%%%%%%%%%%%%%%%%%%%%%%%%%%%%%%%%%%%%%

\subsection{2-loop}

There is one 2-loop topology and three colorings by mass.
One of them reduces to simpler cases,
while the other two are master integrals.
One of the two master integrals has an analytic solution
in terms of Gamma functions.
The other (fully massive) one can be written in terms
of the hypergeometric function $_2F_1$
(see \eqs (4.12) and (4.13) in \re\cite{Davydychev:1992mt}),
or alternatively in terms of a one-dimensional integral
(see \eqs (21), (15) and (16) in \re\cite{Davydychev:1999mq})
which has a simple $\e$-expansion (for 4d in terms of log-sine integrals).
\ba
\twoSa &=& -\frac{d-2}{2(d-3)}\(\oneJ\)^2 \\
\frac{\twoSb}{J^2} &=& \frac{\Gamma(\frac{3-d}2)\Gamma(\frac{d}2)}
 {2^{d-2}\Gamma(\frac12)} \\
\frac{\twoS}{J^2} &=& -\frac{3(d-2)}{4(d-3)}\lb_2F_1\(\frac{4-d}2,1;
 \frac{5-d}2;\frac34\)-3^{\frac{d-5}2}\frac{2\pi\Gamma(5-d)}
 {\Gamma(\frac{4-d}2)\Gamma(\frac{6-d}2)}\rb \\
&=& -\frac{3(d-2)}{4(d-3)}\lb 1-3^{\frac{d-3}2}(d-4)\int_0^{\frac\pi3}d\tau
 (2\sin(\tau))^{4-d}-3^{\frac{d-5}2}\frac{2\pi\Gamma(5-d)}
 {\Gamma(\frac{4-d}2)\Gamma(\frac{6-d}2)}\rb \\
&\stackrel{d=n-2\e}=&
 -\frac{3(n\!-\!2\!-\!2\e)}{4(n\!-\!3\!-\!2\e)}
 \lb 1\!+\!3^{-\e} \frac{\frac{n-4}2-\e}
 {3^{\frac{3-n}2}}\sum_{j=0}^\infty \frac{(2\e)^j}{j!} \ls_{j+1}^{(4-n)}
 \!-\!3^{-\e} \frac{3^{\frac{n-5}2} 2\pi\Gamma(5-n+2\e)}
 {\Gamma(\frac{4-n}2+\e)\Gamma(\frac{6-n}2+\e)}\rb
\ea
The numbers
$\ls_j^{(a)}=-\int_0^{\frac{2\pi}3} d\tau (2\sin\frac\tau2)^a
\ln^{j-1}|2\sin\frac\tau2|$
in the 4d ($n=4$) case are the log-sine integrals
$\ls_j^{(0)}=\ls_j=\ls_j(\frac{2\pi}3)$ of \eq\nr{eq:ls}.

%%%%%%%%%%%%%%%%%%%%%%%%%%%%%%%%%%%%%%%%%%%%%%%%%%%%%%%%%%%%
%%%%%%%%%%%%%%%%%%%%%%%%%%%%%%%%%%%%%%%%%%%%%%%%%%%%%%%%%%%%

\subsection{3-loop}

There are three 3-loop topologies.

%%%%%%%%%%%%%%%%%%%%%%%%%%%%%%%%%%%%%%%%%%%%%%%%%%%%%%%%%%%%

{\bf 3-loop, 4 lines:}
There are four colorings by mass, all of which are master integrals.
Two of them have an analytic solution in terms of Gamma functions.
The third one (called $D_3(0,1,0,1,1,1)$ in the literature,
according to the notation introduced in \re\cite{Avdeev:1995eu})
can be written in terms of a single hypergeometric function $_3F_2$
(see \eq(4.33) of \re\cite{Davydychev:2000na}, where also the first seven
orders of its 4d $\e$-expansion were given in \eq(4.32))\footnote{
In some sense, the representation in terms of special types
of hypergeometric functions can be called an all-order analytic
$\e$\/-expansion, namely when their expansion can be written
in terms of rapidly converging (multiple inverse binomial) sums,
for which efficient algorithms exist \cite{Kalmykov:2000qe}.
The 3-loop integrals $E_3$, $D_5$ and $D_4$ \cite{Davydychev:2003mv}
below belong to this class as well.}.
The first seven orders of the 4d $\e$-expansion of the fourth (fully massive)
master (called $B_N(0,0,1,1,1,1)$ in the literature)
can be deduced from the function $B_4$ introduced in
\re\cite{Broadhurst:1991fi} using the reductions
\eqs\nr{eq:Mb} and \nr{eq:Md} given below.
Two more orders could be obtained from $B_4$ as given in
\re\cite{Broadhurst:1996az}, but we refrain from reproducing
them here.
\ba
\frac{\threeBc}{J^3} &=& -\frac{3\Gamma(\frac{6-3d}2)\Gamma(3-d)
\Gamma^2(\frac{d-2}2)}{\Gamma^3(\frac{2-d}2)} \\
\frac{\threeBa}{J^3} &=& \frac{2^{d-3}\Gamma(\frac{8-3d}2)
\Gamma(\frac{3-d}2)\Gamma(\frac{d}2)}{\Gamma(\frac{7-2d}2)
\Gamma(\frac{2-d}2)} \\
\frac{\threeBb}{J^3}
&\stackrel{d=4-2\e}=&
-1-\frac34\e+\frac18\e^2
+\(\frac{91}{16}-\frac92\sqrt{3}\ls_2\)\e^3
+\(\frac{913}{32}-\frac34\sqrt{3}(\pi^3+9(3\!-\!2\ln3)\ls_2+\!18\ls_3)\)\e^4
\nn&&{}
+\(\frac{7027}{64}+\frac18\sqrt{3}(9\pi^3(2\ln3-3)+64\ls_4(\frac{\pi}3)
-9(67-54\ln3+18\ln^23)\ls_2
\rd\nn&&\ld\vphantom{\frac12}
+162(2\ln3-3)\ls_3-216\ls_4+184\pi\zeta_3)\)\e^5
\nn&&{}
+\(\frac{48601}{128}+\sqrt{3}\(
-\frac3{16}\pi^3(67+18\ln3(\ln3-3))
-\frac{23}{36}\pi^5
+\frac{69}2\pi(3-2\ln3)\zeta_3
+81\pi\ls_4^\prime
\rd\rd\nn&&\ld\vphantom{\frac12}\ld
+\frac9{16}(-457+6\ln3(67+3\ln3(2\ln3-9)))\ls_2
-\frac{27}8(67+8\pi^2+18\ln3(\ln3-3))\ls_3
\rd\rd\nn&&\ld\vphantom{\frac12}\ld
+\frac{81}2(2\ln3-3)\ls_4
-\frac{81}2\ls_5
+12(3-2\ln3)\ls_4(\frac\pi3)
+49\ls_5(\frac\pi3)
-\frac{243}4\ls_5^\prime
\)\)\e^6
\nn&&{}+\order{\e^7}\\
\frac{\threeB}{J^3} &\stackrel{d=4-2\e}=&
-2-\frac53\e-\frac12\e^2+\frac{103}{12}\e^3+\frac7{24}(163-128\zeta_3)\e^4
\nn&&{}+\(\frac{9055}{48}+\frac{136\pi^4}{45}+\frac{32}3\ln^22(\pi^2-\ln^22)
-168\zeta_3-256a_4\)\e^5
\nn&&{}+\(\frac{63517}{96}+\frac{16}5\ln^42(4\ln2-15)
-\frac{16}3\pi^2\ln^22(4\ln2-9)-\frac{68}{15}\pi^4(4\ln2-3)
\rd\nn&&\ld{}-\frac{1876}3\zeta_3+1240\zeta_5-1152a_4-1536a_5\)\e^6
+\order{\e^7}
\ea
Here, $\ls_j^\prime=-\int_0^{\frac{2\pi}3}d\tau \tau^{j-3}
\ln^2|2\sin\frac\tau2|$ are special
values of the generalized log-sine function \cite{Davydychev:2000na}.

%%%%%%%%%%%%%%%%%%%%%%%%%%%%%%%%%%%%%%%%%%%%%%%%%%%%%%%%%%%%

{\bf 3-loop, 5 lines:}
There are eleven colorings by mass.
Eight of them reduce, while the remaining three are master integrals.
One of the masters has an analytic solution in terms of Gamma functions.
The second one (called $E_3$ in the literature) can be written in terms
of the hypergeometric function $_2F_1$,
cf. \eq(4.24) of \re\cite{Davydychev:2000na}.
Its first six terms of the 4d $\e$-expansion (we will only reproduce the first
five of them below) are given
in \eqs(4.16),(4.18) of \re\cite{Davydychev:2000na}.
The first five terms of the 4d $\e$-expansion of the third (fully massive)
master integral (called $D_5$ in the literature)
can be deduced from \re\cite{Broadhurst:1998rz}
using the reduction given in \eq\nr{eq:Ma} below.
One more term has recently been given in 
\eq(3.28) of \re\cite{Kalmykov:2005hb},
but we refrain from listing it here.
\ba
\threeVa &=& \frac1{6(d-3)}\lb(3d-8)\threeBb-3(d-2)\twoS\oneJ\rb\\
\threeVc &=& \frac1{4(d-3)}\lb(3d-8)\threeBa
 +\frac{(d-2)^2}{d-3}\(\oneJ\)^3\rb \\
\threeVd &=& \frac1{2(d-3)}\lb(3d-8)\threeBc-(d-2)\twoSb\oneJ\rb\\
\threeVf &=& -\frac1{4(d-4)}\lb(3d-8)\threeB+\frac{2(d-2)^2}{d-3}
\(\oneJ\)^3\rb\\
\threeVg &=& \frac{3d-8}2\threeBb-\frac{d-2}2\twoSb\oneJ
+\frac{(d-2)^2}{2(d-3)}\(\oneJ\)^3\\
\threeVh &=& -\frac{3d-8}{4(2d-7)}\threeBa \\
\threeVi &=& \frac{3d-8}{d-2}\threeBa-2\twoSb\oneJ \\
\threeVj &=& -\frac{3d-8}{d-4}\threeBc\\
\frac{\threeVe}{J^3} &=& \frac{\pi^3}{\sin^2(\frac{\pi d}2)
\sin(\frac{3\pi d}2)} \frac{\Gamma^2(\frac{d-2}2)}
{\Gamma^3(\frac{2-d}2)\Gamma^2(d-2)\Gamma(\frac{d}2)}\\
\frac{\threeVb}{J^3} &\stackrel{d=4-2\e}=&
\frac23+\frac53\e
+\(5+\frac{\pi^2}6-3\sqrt{3}\ls_2\)\e^2
\nn&&{}
+\(\frac{44}3+\frac{\pi^2}3+\frac13\zeta_3-3\sqrt{3}\(\frac{5\pi^3}{162}
+(2-\ln3)\ls_2+\ls_3\)\)\e^3
\nn&&{}
+\(\frac{128}3+\frac{5\pi^2}6-\frac{\pi^4}{60}+\frac{10}3\zeta_3
+\sqrt{3}\(
-\frac16(2\pi^2+9\ln3(10+(\ln3-4)\ln3))\ls_2
\rd\rd\nn&&\ld\ld
+3(\ln3-2)\ls_3-2\ls_4-\frac{80}{27}\ls_4(\frac\pi3)
+\frac{5\pi^3}{54}(\ln3-2)+\frac{94}{27}\pi\zeta_3\)
\)\e^4
+\order{\e^5}\\
\frac{\threeV}{J^3} &\stackrel{d=4-2\e}=&
1+\frac83\e+\(\frac{25}3-6\sqrt{3}\ls_2\)\e^2+\(\frac{76}3-6\zeta_3
+\sqrt{3}\(-\frac{\pi^3}3+6(\ln3-2)\ls_2-6\ls_3\)\)\e^3
\nn&&{}+\(76-\frac{7\pi^4}{10}+18\ls_2^2-12\pi\ls_3+18\ls_4^\prime
+\(-\frac{92}3+4\sqrt{3}\pi+26\ln3\)\zeta_3
\rd\nn&&\ld+\sqrt{3}\(\frac{\pi^3}3(\ln3\!-\!2)
\!-\!3(10\!-\!4\ln3\!+\!\ln^23)\ls_2
\!+\!6(\ln3\!-\!2)\ls_3\!-\!4\ls_4\)\)\e^4
\!+\!\order{\e^5}
\ea

%%%%%%%%%%%%%%%%%%%%%%%%%%%%%%%%%%%%%%%%%%%%%%%%%%%%%%%%%%%%

{\bf 3-loop, 6 lines:}
There are ten colorings by mass.
The first two terms of their 4d $\e$-expansion are given in
\re\cite{Broadhurst:1998rz}.
Five of them reduce, while the remaining five are masters.
The third term of the 4d $\e$-expansion of one of the masters
(called $D_4$ in the literature) is given in \eq(4.10) of
\re\cite{Davydychev:2000na}\footnote{Note that there is a typo
in \eq(4.10) of \re\cite{Davydychev:2000na}. The second-last term should
read $-\frac{161}{54}\pi \ls_4(\frac{\pi}3)$, see also
\re\cite{Kalmykov:2005hb}.}.
The remaining four masters (called $D_M$, $D_N$, $D_3$ and $D_6$,
respectively) are read from \re\cite{Broadhurst:1998rz}.
\ba
\threeMa &=& \label{eq:Ma}
-\frac{2(d-3)}{3(d-4)}\threeV +\frac{3d-8}{12(d-4)}\threeB
-\frac{2(d-2)}{3(d-4)}\twoS\oneJ
\nn&&{}-\frac{(d-2)^2}{6(d-4)(d-3)}\(\oneJ\)^3 \\
\threeMb &=& \label{eq:Mb} \frac{(3d\!-\!10)(3d\!-\!8)}{16(d-4)^2}
\(\threeB+\frac{4(d\!-\!4)}{2d-7}\threeBa\)
+\frac{(d\!-\!2)^2(5d\!-\!18)}{8(d-4)^2(d-3)}
\(\oneJ\)^3\\
\threeMd &=& \label{eq:Md} -\frac{3(3d-10)(3d-8)}{16(d-4)(2d-7)}\threeBa 
-\frac{(d-2)^2}{8(d-4)(d-3)}\(\oneJ\)^3\\
\threeMg &=& -\frac{(3d-10)(3d-8)}{(d-4)^2}\(\threeBc
+\frac{d-4}{4(2d-7)}\threeBa\)+\frac{d-2}{d-4}\twoSb\oneJ \\
\threeMi &=& \frac{2(d-3)}{d-4}\threeVe 
+\frac{2(3d-10)(3d-8)}{(d-4)^2}\threeBc \\
\frac{\threeMc}{J^3} &\stackrel{d=4-2\e}=&
-2\zeta_3\e^2+\(\frac{77\pi^4}{1080}+\frac{27}2\ls_2^2\)\e^3
\nn&&{}+\(-\frac{21}8\chi_5+\frac{161}{54}\pi
\ls_4(\frac{\pi}3)-\frac{367}{216}\pi^3 \ls_2-7\pi
\ls_4-2\zeta_3+\frac{2615}{432}\pi^2 \zeta_3-\frac{2047}{216}\zeta_5\)\e^4
\nn&&{}+\order{\e^5} \\
\frac{\threeMf}{J^3} &\stackrel{d=4-2\e}=& -2\zeta_3\e^2
+\(\frac{11\pi^4}{180}+9\ls_2^2\)\e^3+\order{\e^4}\\
\frac{\threeMh}{J^3} &\stackrel{d=4-2\e}=&
-2\zeta_3\e^2+\(\frac{7\pi^4}{60}+\frac23\ln^22(\pi^2-\ln^22)-16a_4\)\e^3
+\order{\e^4}\label{eq:MhAnal}\\
\frac{\threeMe}{J^3} &\stackrel{d=4-2\e}=& -2\zeta_3\e^2+\(\frac{\pi^4}{24}
+\frac{27}2\ls_2^2\)\e^3+\order{\e^4} \\
\frac{\threeM}{J^3} &\stackrel{d=4-2\e}=& -2\zeta_3\e^2+\(\frac{17\pi^4}{90}
+\frac23\ln^22(\pi^2-\ln^22)+9\ls_2^2-16a_4\)\e^3+\order{\e^4}
\ea

Here, $\chi_5=\sum_{n=1}^{\infty}\frac{(n!)^2}{(2n)!}\frac1{n^2}
\sum_{j=1}^{n-1}\frac1j\approx0.0678269619272...$ 
is a special case of a binomial sum \cite{Davydychev:2000na}.

%%%%%%%%%%%%%%%%%%%%%%%%%%%%%%%%%%%%%%%%%%%%%%%%%%%%%%%%%%%%
%%%%%%%%%%%%%%%%%%%%%%%%%%%%%%%%%%%%%%%%%%%%%%%%%%%%%%%%%%%%

\subsection{4-loop}

There are ten topologies.

%%%%%%%%%%%%%%%%%%%%%%%%%%%%%%%%%%%%%%%%%%%%%%%%%%%%%%%%%%%%

{\bf 4-loop QED-type cases, 5 lines:}
There is one topology, BB.

There are two QED-type colorings of BB. Both of them are masters.
One is known analytically in terms of Gamma functions,
while the other one is new. 
Interestingly, the analytic value of the last term in \eq\nr{eq:BB4}
was obtained by a physics computation in which this master integral
contributed \cite{MSYSnew}.
\ba
\frac{\fourBBa}{J^4}
 &=& 3 (d-2) 4^{d-3} \frac{\Gamma(5-2d)\Gamma(\frac{8-3d}2)
\Gamma(\frac{5-d}2)\Gamma^2(\frac{d}2)}{\Gamma(\frac{11-3d}2)
\Gamma^3(\frac{4-d}2)} \\
\frac{\fourBBb}{J^4} &\stackrel{d=4-2\e}=& \label{eq:BB4}
-1-\frac12\e+\frac{17}{36}\e^2+\frac1{216}\e^3-\frac{37207}{1296}\e^4
+\(-\frac{1976975}{7776}+\frac{1792}9\zeta_3\)\e^5+\order{\e^6}
\ea

%%%%%%%%%%%%%%%%%%%%%%%%%%%%%%%%%%%%%%%%%%%%%%%%%%%%%%%%%%%%

{\bf 4-loop QED-type cases, 6 lines:}
There are two topologies, T and G.

There are four QED-type colorings of T. All of them are masters.
One of them is known analytically, while the first six orders of
the 4d $\e$-expansion of two others were given
in \eq(16) of \re\cite{Laporta:2002pg}
and \eq(18) of \re\cite{Chetyrkin:2004fq}\footnote{Note that in
\re\cite{Chetyrkin:2004fq} the last term of \eq\nr{TcChetyrkin}
involves a numerical coefficient
$N_{10}\approx5.3111546$, which we have determined to be
$N_{10}=\fr{49\pi^4}{720}+\fr{1}{6}\ln^2 2\(\pi^2-\ln^2 2\)-4a_4$, using
our high-precision result \eq\nr{Tc}
and PSLQ \cite{pslq}.}, respectively.
The fourth one is new.
\ba
\frac{\fourTa}{J^4} &=&
\frac{8^{d-3}\Gamma^3(\frac12)\Gamma(6-2d)\Gamma^3(\frac{d}2)}
{\sin(\frac{3\pi d}2)\Gamma(\frac{11-3d}2)\Gamma^2(\frac{4-d}2)
\Gamma^2(d-2)} \\
\frac{\fourT}{J^4}
 &\stackrel{d=4-2\e}=& \label{TLaporta}
 \frac32+\frac72\e+\frac92\e^2+\(-\frac{39}2-3\zeta_3\)\e^3
+\(-208+\frac{\pi^4}{20}+109\zeta_3\)\e^4
\nn&&{}+\(-1254-\frac{547\pi^4}{60}+32\ln^22(\ln^22-\pi^2)+768a_4
 +855\zeta_3+189\zeta_5\)\e^5
\nn&&{}+\order{\e^6} \\
\frac{\fourTc}{J^4}
 &\stackrel{d=4-2\e}=& \label{TcChetyrkin}
\frac23+\frac43\e+\frac23\e^2+\frac43(-11+4\zeta_3)\e^3
+\(-116-\frac{4\pi^4}{15}+\frac{200\zeta_3}3\)\e^4
\nn&&{}+\(-\frac{1928}3-\frac{326\pi^4}{45}+\frac{64}3\ln^22(\ln^22-\pi^2)
+512a_4+\frac{1192\zeta_3}3+96\zeta_5\)\e^5
\nn&&{}+\order{\e^6} \\
\frac{\fourTb}{J^4} &\stackrel{d=4-2\e}=&
 \frac14+\frac12\e+0\cdot\e^2+\(-8+\frac{13}2\zeta_3\)\e^3
+\(-\frac{241}4-\frac{5\pi^4}8+4\zeta_3\)\e^4
\nn
&&{}+\(-\fr{669}{2}-\fr{\pi^4}{5}+36\zeta_3+\fr{693}{2}\zeta_5\)\e^5
+\order{\e^6} \label{Tbana} \ea

There are five QED-type colorings of G. All of them reduce.
\ba
\fourGb
  &=& \fr{2d-5}{4(d-3)} \fourBBb -\fr{d-2}{2(d-3)} \threeB \oneJ \\
\fourGc
  &=& \fr{2d-5}{4} \fourBBb -\frac{d-2}2 \threeBa \oneJ
  +\frac{(d-2)^3}{8(d-3)^2} \(\oneJ\)^4\\
\fourGd
  &=& -\fr{2d-5}{6(d-3)} \fourBBa \\
\fourGe
  &=& -\fr{(2d-5)(3d-8)}{6(d-3)(d-4)} \fourBBa \\
\fourGa
  &=& \fr{2d-5}{2(d-3)} \fourBBa -\fr{d-2}{2(d-3)} \threeBa \oneJ
\ea

%%%%%%%%%%%%%%%%%%%%%%%%%%%%%%%%%%%%%%%%%%%%%%%%%%%%%%%%%%%%

{\bf 4-loop QED-type cases, 7 lines:}
There are three topologies, VB, N and U.

There are seven QED-type colorings of VB.
Five of them reduce.
There are two master integrals.
Both are new.
\ba
\fourVBd &=&
 \frac{(2d-5)(3d-10)(3d-8)}{3(d-4)^2(3d-11)}\fourBBa
 +\frac{2(d-3)^2}{(d-4)(3d-11)}\fourTa\\
\fourVBe &=&
 -\frac{(2d-5)(3d-8)}{3(d-4)^2}\fourBBa
 -\frac{4(d-3)^2}{3(d-4)(3d-10)}\fourTb \nn&&{}
 +\frac{(d-2)(3d-8)}{8(2d-7)(3d-10)}\threeBa\oneJ\\
\fourVBf &=&
 \frac{(2d-5)(3d-8)}{3(d-4)(d-3)}\fourBBa
 -\frac{(d-3)}{d-4}\fourTc
 -\frac{(d-2)(3d-8)}{4(d-4)(d-3)}\threeBa\oneJ\\
\fourVBg &=&
 \frac{(2d-5)(3d-8)}{16(d-4)(d-3)}\fourBBb
 -\frac{2(d-3)}{3(d-4)}\fourT
 -\frac{(d-2)(3d-8)}{8(d-4)(d-3)}\threeB\oneJ \nn&&{}
 -\frac{(d-2)^3}{32(d-4)(d-3)^2}\(\oneJ\)^4\\
\fourVBc &=& \frac2{3d\!-\!10}\lb(d\!-\!3)\fourTa
-\frac{3d-8}{2(d\!-\!3)}\(\frac{2d\!-\!5}3
 \fourBBa-\frac{d\!-\!2}4 \threeBa\oneJ\)\rb \\
\frac{\fourVBb}{J^4} &\stackrel{d=4-2\e}=&
-\frac16-\frac56\e-\(\frac{11}3+\zeta_3\)\e^2
+\(-\frac{44}3-\frac{\pi^4}{60}+\frac23\zeta_3\)\e^3
\nn&&{}+\(-\frac{332}6-\frac{\pi^4}6+\frac{31}3\zeta_3+53\zeta_5\)\e^4
+\order{\e^5}\\
\frac{\fourVBa}{J^4} &\stackrel{d=4-2\e}=&
-\frac16-\frac56\e-\(\frac{11}3+\frac12\zeta_3\)\e^2
+\(-\frac{44}3-\frac{\pi^4}{120}+\frac{13}6\zeta_3\)\e^3
\nn&&{}+\(-\frac{166}3-\frac{5\pi^4}{24}+\frac{29}6\zeta_3
+\frac{43}2\zeta_5\)\e^4+\order{\e^5}
\ea

There are five QED-type colorings of N. All of them reduce.
\ba
\fourNa
 &=& \frac{(2d-5)(3d-8)}{6(d-3)^2}\fourBBa-\frac{(d-2)(3d-8)}{4(d-3)^2}
\threeBa\oneJ-\frac{(d-2)^3}{8(d-3)^3}\(\oneJ\)^4 \\
\fourNb
  &=& -\fr{(2d-5)(3d-8)}{6(d-4)(d-3)} \fourBBa 
+\fr{(d-2)(3d-8)}{8(d-3)(2d-7)}\threeBa\oneJ\\
\fourNc
  &=& \fr{(2d-5)(3d-8)}{6(d-4)(3d-11)}\fourBBa\\
\fourNd
 &=& -\frac{(2d-5)(3d-8)}{16(d-4)}\fourBBb
+\frac{(d-2)(3d-8)}{4(2d-7)}\threeBa\oneJ
\nn&&{}-\frac{3(d-2)^3}{32(d-4)(d-3)}\(\oneJ\)^4\\
\fourNe
 &=&
\frac{(2d-5)(3d-8)}{16(d-3)}\fourBBb+\frac{(d-2)(3d-8)}{8(d-4)(d-3)}
\threeB\oneJ
\nn&&{}-\frac{(d-2)(3d-8)}{8(d-3)}\threeBa\oneJ
+\frac{(d-2)^3(3d-4)}{32(d-4)(d-3)^2}\(\oneJ\)^4
\ea

There are four QED-type colorings of U. All of them reduce.
\ba
\fourUa
  &=& \fr23\fourTb -\fr{(d-2)(3d-8)}{8(d-3)^2}\threeBa\oneJ
  -\fr{(d-2)^3}{8(d-3)^3}\(\oneJ\)^4 \\
\fourUb
  &=&  -\fr{d-3}{d-5}\fourT -\frac{3(d-2)(3d-8)}{8(d-5)(d-4)}\threeB\oneJ
  -\frac{3(d-2)^3}{4(d\!-\!5)(d\!-\!4)(d\!-\!3)}\(\oneJ\)^4\\
\fourUc
  &=& -\fr{d-3}{2(d-4)}\fourTc -\fr{(d-2)(3d-8)}{8(d-4)(2d-7)}\threeBa\oneJ\\
\fourUd
  &=& -\fr{d-3}{3d-11}\fourTa
\ea

%%%%%%%%%%%%%%%%%%%%%%%%%%%%%%%%%%%%%%%%%%%%%%%%%%%%%%%%%%%%

{\bf 4-loop QED-type cases, 8 lines:}
There are two topologies, VV and W.

There are seven QED-type colorings of VV. All of them reduce.
\ba
\fourVVa &=& -\fr{(2d-7)(2d-5)(3d-10)(3d-8)}{6(d-4)^2(3d-13)(3d-11)}\fourBBa
-\fr{(d-3)^2(2d-7)}{(d-4)(3d-13)(3d-11)}\fourTa
\\
\fourVVb &=& \fr{(2d-7)(2d-5)(3d-10)(3d-8)}{18(d-4)^2(d-3)(3d-11)}\fourBBa
- \fr{(d-3)(2d-7)}{3(d-4)(3d-11)}\fourTa \nn
&&{}-\fr{(d-2)(3d-10)(3d-8)}{48(d-4)(d-3)(2d-7)}\threeBa\oneJ\\
\fourVVc &=& -\fr{(2d-7)(2d-5)(3d-10)(3d-8)}{6(d-4)^2(d-3)(3d-11)}\fourBBa
+ \fr{(d-3)(2d-7)}{4(d-4)^2}\fourTc \nn
&&{}+\fr{(d-2)(3d-8)(5d^2-35d+61)}{16(d-4)^2(d-3)(2d-7)}\threeBa\oneJ\\
\fourVVd &=& -\fr{(2d-7)(2d-5)(3d-10)(3d-8)}{6(d-4)^2(d-3)(3d-11)}\fourBBa
- \fr{4(d-3)^2(2d-7)}{3(d-4)(3d-11)(3d-10)}\fourTb\\
&&{}+ \fr{(d-2)(3d-8)(95d^3-989d^2+3428d-3956)}
{32(d-4)(d-3)(2d-7)(3d-11)(3d-10)}\threeBa\oneJ
\fr{(d-2)^3}{16(d-4)(d-3)^2}\(\oneJ\)^4\nn
\fourVVe &=& -\fr{(2d-5)(3d-11)(3d-8)}{32(d-4)(2d-9)}\fourBBb
- \fr{(d-3)^2}{4(d-4)(2d-9)}\fourTc -
 \fr{(3d-10)}{4(2d-9)}\fourVBb \nn
&&{}+ \fr{(d-2)(3d-8)(12d^2-101d+204)}{32(d-4)(2d-9)(2d-7)}\threeBa\oneJ\nn
 &&{}- \fr{(d-2)(3d-8)}{16(d-4)(2d-9)}\threeB\oneJ
 - \fr{3(d-2)^3(3d-7)}{64(d-4)(d-3)(2d-9)}\(\oneJ\)^4
\\
\fourVVf &=& \fr{(2d-5)(3d-8)}{12(d-3)(2d-7)}\fourBBa +
 \fr{(2d-5)(3d-8)}{64(d-3)(2d-7)}\fourBBb -
 \fr{2(d-3)^2}{3(2d-7)(3d-10)}\fourTb \nn
 &&{}+
 \fr{3d-10}{2(2d-7)}\fourVBa -
 \fr{(d-2)^3(73d^2-512d+896)}{128(d-4)^2(d-3)^2(2d-7)}\(\oneJ\)^4\nn
&&{} -  \fr{(d-2)(3d-8)(19d^2-128d+216)}{(16(d-4)(d-3)(2d-7)(3d-10)}
   \threeBa\oneJ \\
&&{}
-\fr{(d-2)(3d-10)(3d-8)}{32(d-4)^2(d-3)}\threeB\oneJ
- \fr{(d-2)^3(73d^2-512d+896)}{128(d-4)^2(d-3)^2(2d-7)}\(\oneJ\)^4\nn
\fourVVg &=& \fr{(2d-7)(2d-5)(3d-10)(3d-8)}{64(d-4)^2(d-3)}\fourBBb +
\fr{(d-3)(2d-7)}{3(d-5)(d-4)}\fourT \nn
&&{} -
 \fr{(d\!-\!2)(3d\!-\!10)(3d\!-\!8)}{8(d-4)(2d-7)}\threeBa\oneJ
+ \fr{(d-2)(3d-8)(11d^2-77d+134)}{32(d-5)(d-4)^2(d-3)}\threeB\oneJ\nn
&&{}+\fr{(d-2)^3(18d^3-129d^2+245d-58}{128(d-5)(d-4)^2(d-3)^2}\(\oneJ\)^4
\ea

There are five QED-type colorings of W.
Four of them reduce.
There is one master integral, which is new.
\ba
\fourWd &=& -\fr{(2d-5)(3d-8)(27d^3-283d^2+990d-1156)}
{12(d-4)^3(2d-7)(3d-11)}\fourBBa
- \fr{(d-3)(3d-10)}{2(d-4)(2d-7)}\fourVBa \nn
&&{}- \fr{2(d-3)^4(5d-18)}{3(d-4)^2(2d-7)(3d-11)(3d-10)}\fourTb
- \fr{(2d-5)(3d-8)}{64(d-4)(2d-7)}\fourBBb \\
&&{}+ \fr{(d-2)(3d-8)(5d-18)}{16(d-4)(3d-11)(3d-10)}\threeBa\oneJ
 - \fr{7(d-2)^3}{128(d-4)(d-3)(2d-7)}\(\oneJ\)^4 \nn
\fourWe &=& \fr{2(2d-7)(2d-5)(3d-10)(3d-8)}{9(d-4)^3(3d-11)}\fourBBa +
\fr{8(d-3)^3(2d-7)}{9(d-4)^2(3d-11)(3d-10)}\fourTb \nn
&&{}+ \fr{2(d-3)^3(2d-7)}{3(d-4)^2(3d-11)}\fourTa
-\fr{(d-2)(3d-8)(7d^2-48d+82)}{24(d\!-\!4)(2d\!-\!7)(3d\!-\!11)(3d\!-\!10)}
\threeBa\oneJ \\
\fourWb &=&\fr{(2d-5)(3d-8)}{6(d-4)(2d-7)}\fourBBa
-\fr{(2d-5)(3d-8)}{32(d-3)(2d-7)}\fourBBb+\fr{d-3}{2(d-4)}\fourTc\nn
&&{}-\fr{4(d-3)^3}{3(d-4)(2d-7)(3d-10)}\fourTb
+\fr{(d-3)(3d-10)}{(d-4)(2d-7)}\fourVBa
-\fr{3d-10}{2(d-4)}\fourVBb\nn
&&{}-\fr{(d-2)(3d-8)}{8(2d-7)(3d-10)}\threeBa\oneJ
+\fr{(d-2)(3d-8)}{16(d-4)(d-3)}\threeB\oneJ\nn
&&{}+\fr{(d-2)^3(9d-28)}{64(d-4)(d-3)^2(2d-7)}\(\oneJ\)^4 \\
\fourWc &=& + \fr{(2d-5)(3d-8)}{32(d-4)(d-3)}\fourBBb
+ \fr{2(d-3)^2(2d-7)}{3(d-4)^2(3d-11)}\fourT -\fr{d-3}{2(d-4)}\fourTc \nn
&&{}+\fr{3d-10}{2(d-4)}\fourVBb
 + \fr{(d-2)(3d-8)(d^2-4d+2)}{8(d-4)^2(d-3)(3d-11)}\threeB\oneJ\nn
&&{}+ \fr{(d-2)^3(29d^2-177d+268)}{64(d-4)^2(d-3)^2(3d-11)}\(\oneJ\)^4
 \\
\frac{\fourWa}{J^4} &\stackrel{d=4-2\e}=& 5\zeta_5\e^3+\order{\e^4}
\ea

%%%%%%%%%%%%%%%%%%%%%%%%%%%%%%%%%%%%%%%%%%%%%%%%%%%%%%%%%%%%

{\bf 4-loop QED-type cases, 9 lines:}
There are two topologies, H and X.

There are five QED-type colorings of H. All of them reduce.
\ba
\fourHa &=& -\fr{3(2d-5)(3d-11)(3d-8)(54 - 29d + 4d^2)}
{256(d-5)(d-4)^2(d-3)(2d-9)}\fourBBb
- \fr{4(d-3)^2(2d-7)}{3(d-5)(3d-13)(3d-11)}\fourT \nn
&&{}+
\fr{3(d-3)(d^2-11d+27)}{8(d-5)(d-4)(2d-9)}\fourTc
- \fr{9(d-4)(3d-10)}{8(d-5)(2d-9)}\fourVBb \nn
&&{}-\fr{(d-2)(3d-8)(13d^3-77d^2+9d+351)}
{32(d-5)(d-3)(2d-9)(3d-13)(3d-11)}\threeB\oneJ \nn
&&{}-\fr{3(d-2)(3d-8)(13d^2-92d+162)}{64(d-5)(d-4)(2d-9)(2d-7)}\threeBa\oneJ\\
&&{}- \fr{(d-2)^3(2452d^5-43031d^4+329345d^3-1198763d^2+2170827d-1564110)}
{512(d-5)(d-4)^2(d-3)^2(2d-9)(3d-13)(3d-11)}\(\oneJ\)^4 \nn
\fourHb &=& -\fr{3(2d-7)(2d-5)(3d-10)(3d-8)}{2(d-4)(3d-14)(3d-13)(3d-11)}
\fourBBa - \fr{6(d-3)^2(2d-7)}{(3d-14)(3d-13)(3d-11)}\fourTa \nn
&&{}- \fr{32(d-3)^3(2d-7)}{9(3d-14)(3d-13)(3d-11)(3d-10)}\fourTb \nn
&&{}+
\fr{(d-2)(3d-8)(139d^3-1495d^2+5344d-6348)}
{96(2d-7)(3d-14)(3d-13)(3d-11)(3d-10)}\threeBa\oneJ
\\
\fourHc &=& \fr{7(2d-7)(2d-5)(3d-10)(3d-8)}{9(d-4)^2(3d-13)(3d-11)}\fourBBa
+ \fr{4(d-3)^2(2d-7)}{3(d-4)(3d-13)(3d-11)}\fourTa \nn
&&{}+\fr{32(d-3)^3(2d-7)}{9(d-4)(3d-13)(3d-11)(3d-10)}\fourTb -
\fr{d-4}{2(3d-13)}\fourWa \nn
&&{}-
\fr{(d-2)(3d-8)(409d^3-4285d^2+14944d-17348)}
{96(d-4)(2d-7)(3d-13)(3d-11)(3d-10)}\threeBa\oneJ\nn
&&{} - \fr{(d-2)^3}{16(d-4)(d-3)(3d-13)}\(\oneJ\)^4
\\
\fourHd &=& -\fr{(2d-5)(3d-8)(3d^3-19d^2+19d+38)}
{12(d-5)(d-4)^2(2d-7)(3d-11)}\fourBBa \nn
&&{} +
\fr{(2d-5)(3d-11)(3d-8)(2d^3-14d^2+23d+6)}{128(d-5)(d-4)^3(d-3)(2d-7)}
\fourBBb \nn
&&{}-\fr{2(d-3)^2(2d-7)}{3(d-5)(d-4)(3d-11)}\fourT -
\fr{2(d-3)^3(17d^2-125d+230)}{3(d-5)(d-4)(2d-7)(3d-11)(3d-10)}\fourTb \nn
&&{}+ \fr{(d-3)(5d^2-37d+69)}{4(d-5)(d-4)^2}\fourTc +
\fr{3(d-3)(3d-10)}{2(d-5)(2d-7)}\fourVBa - \fr{3(3d-10)}{4(d-5)}\fourVBb  \nn
&&{}-
\fr{(d-2)(3d-8)(10d^3-87d^2+235d-186)}{32(d-5)(d-4)^2(d-3)(3d-11)}
\threeB\oneJ \nn
&&{}-
\fr{(d-2)(3d-8)(38d^4-568d^3+3174d^2-7863d+7290)}
{16(d-5)(d-4)^2(2d-7)(3d-11)(3d-10)}
\threeB\oneJ\nn
&&{} -
\fr{(d-2)^3(418d^5-7346d^4+51389d^3-178846d^2+309603d-213234)}
{256(d-5)(d-4)^3(d-3)^2(2d-7)
(3d-11)}\(\oneJ\)^4 \\
\fourHe &=& -\fr{(2d-5)(3d-8)(69d^3-725d^2-2543d-2978)}
{12(d-4)^2(2d-9)(2d-7)(3d-11)}\fourBBa \nn
&&{}-\fr{3(2d-5)(3d-8)}{128(d-3)(2d-9)(2d-7)}\fourBBb -
\fr{2(d-3)^3(d-2)(13d-47)}{9(d-4)(2d-9)(2d-7)(3d-11)(3d-10)}\fourTb \nn
&&{}- \fr{(d-3)(d^2-11d+27)}{8(d-4)^2(2d-9)}\fourTc 
- \fr{3(d-3)(3d-10)}{2(2d-9)(2d-7)}\fourVBa \nn
&&{}+\fr{3(3d-10)}{8(2d-9)}\fourVBb 
+\fr{(d-2)(3d-8)(5d-18)}{64(d-4)^2(d-3)(2d-9)}\threeB\oneJ\nn
&&{} +\fr{(d-2)(3d-8)(1139d^4-16453d^3+89068d^2-214178+193044)}
{192(d-4)^2(2d-9)(2d-7)(3d-11)(3d-10)}\threeBa\oneJ\nn
&&{} +
\fr{(d-2)^3(48d^3-445d^2+1352d-1344)}{256(d-4)^2(d-3)^2(2d-9)(2d-7)}
\(\oneJ\)^4
\ea

There are two QED-type colorings of X.
One reduces.
The other one is a master integral.
It is new.
\ba
\fourXb &=& \fr{(2d-5)(3d-8)(2109d^4-31288d^3+173302d^2-425005d+389562)}
{36(d-4)(2d-9)^2(2d-7)(3d-13)(3d-11)}\fourBBa \nn
&&{}+
\fr{(2d-5)(3d-8)(24d^3-268d^2+1003d-1258)}{256(d-4)^2(2d-9)^2(2d-7)}
\fourBBb \nn
&&{}+
\fr{14(d-3)^2(2d-7)}{3(2d-9)(3d-13)(3d-11)}\fourTa \nn
&&{}+
\fr{2(d-3)^3(295d^3-3332d^2+12431d-15334)}
{9(2d-9)^2(2d-7)(3d-13)(3d-11)(3d-10)}\fourTb \nn
&&{}+
\fr{3(d-4)(d-3)(3d-10)}{2(2d-9)^2(2d-7)}\fourVBa +
\fr{(d-4)^2}{2(2d-9)(3d-13)}\fourWa\nn
&&{}-
\fr{(d-2)(3d-8)(599d^4-9067d^3+51340d^2-12886d+121044)}
{48(2d-9)^2(2d-7)(3d-13)(3d-11)
(3d-10)}\threeBa\oneJ\nn
&&{}+
\fr{(d-2)^3(392d^4-6204d^3+36843d^2-97323d+96502)}
{512(d-4)^2(d-3)(2d-9)^2(2d-7)(3d-13)}\(\oneJ\)^4
\\
\frac{\fourXa}{J^4} &\stackrel{d=4-2\e}=& X_0 \e^4+\order{\e^5}
\ea

All the above formulas agree with our numerical results
of Section~\ref{se:four}.

%%%%%%%%%%%%%%%%%%%%%%%%%%%%%%%%%%%%%%%%%%%%%%%%%%%%%%%%%%%%
%%%%%%%%%%%%%%%%%%%%%%%%%%%%%%%%%%%%%%%%%%%%%%%%%%%%%%%%%%%%
%%%%%%%%%%%%%%%%%%%%%%%%%%%%%%%%%%%%%%%%%%%%%%%%%%%%%%%%%%%%

\section{Conclusions}

We have employed the general method of numerically solving 
single-scale integrals in terms of their $\e$\/-expansion
around $d=4-2\e$ via difference equations, to high precision
and to high $\e$\/-orders. 
We have covered the set of all vacuum master integrals up to three
loops, as well as `QED-type' vacuum master integrals at 4-loop
order. 
These integrals play a role in state-of-the-art perturbative
calculations for precision tests of the standard model.

The main vehicle of solving the difference equations treated
in this work was a formal representation in terms of factorial 
series, which could then be evaluated numerically in a 
truncated form.

In cases where the factorial series representation does not
converge, a more general (and hence more complicated) method
can be used, which transforms the problem into differential
equations. We have encountered only one such case, and have 
shown in detail how it can be represented in terms of multiple
integrals, which we then solved numerically.

Furthermore, we have made an attempt to collect all presently
known analytic results for the class of vacuum master integrals
that we have treated here, up to the 4-loop level. 
This is meant as a concise reference for practitioners in the field.

%%%%%%%%%%%%%%%%%%%%%%%%%%%%%%%%%%%%%%%%%%%%%%%%%%%%%%%%%%%%
%%%%%%%%%%%%%%%%%%%%%%%%%%%%%%%%%%%%%%%%%%%%%%%%%%%%%%%%%%%%
%%%%%%%%%%%%%%%%%%%%%%%%%%%%%%%%%%%%%%%%%%%%%%%%%%%%%%%%%%%%

\section*{Acknowledgments}

We thank M.~Kalmykov for comments on the 3-loop integrals,
S.~Moch for correspondence on harmonic sums,
and M.~Steinhauser for reading the manuscript.
Y.S. would like to thank J.M.~Gelinas and the MIT/LNS
computer services group for their efforts with
installing Condor \cite{condor}, which helped finding
the difference equations needed for this work in finite time.
A.V. was supported in part by the Foundation of Magnus Ehrnrooth.

%%%%%%%%%%%%%%%%%%%%%%%%%%%%%%%%%%%%%%%%%%%%%%%%%%%%%%%%%%%%
%%%%%%%%%%%%%%%%%%%%%%%%%%%%%%%%%%%%%%%%%%%%%%%%%%%%%%%%%%%%
%%%%%%%%%%%%%%%%%%%%%%%%%%%%%%%%%%%%%%%%%%%%%%%%%%%%%%%%%%%%

\begin{appendix}

\section{Numerical results for analytically known master integrals}

As a complement to Section~\ref{se:four}, we here 
list the first few terms of the 
Laurent expansions in $\e=(4-d)/2$ of those
single-mass-scale vacuum master integrals up to four loops that are
known analytically (see the explicit $d$\/-dimensional 
expressions of Section~\ref{se:Analytic}).

Notation and integral measure are as in Section~\ref{se:four},
which in particular determines the 1-loop tadpole to be
$J=\fr{1}{\Gamma(2+\e)}\int\!\fr{{\rm d}^{4-2\e}p}{\pi^{2-\e}}\,
\frac1{p^2+1}=\frac{-1}{\e(1-\e^2)}=-\sum_{n=0}^\infty \e^{2n-1}$.

\ba
\oneJ &=&{} \label{J}
-1.0000000000000000000000000000000000000000000000000 \ep^{-1}
\nN&&{}-1.0000000000000000000000000000000000000000000000000 \ep
\nN&&{}-1.0000000000000000000000000000000000000000000000000 \ep^3
       +\order{\e^{5}}\\
\twoSb &=&{} \label{Sb}
-0.5000000000000000000000000000000000000000000000000\ep^{-2}
\nN&&{}-0.5000000000000000000000000000000000000000000000000 \ep^{-1}
\nn&&{}-3.6449340668482264364724151666460251892189499012068
\nn&&{}-3.4428771636886321510726770051345751984539636088663 \ep
\nn&&{}-17.748133915933433322311939139507906328597192596121 \ep^2
\nn&&{}-16.366439374126401323669287645924253086404587829687 \ep^3
       +\order{\e^{4}}\\
%\nn&&{}-75.837518318238099881147681234139872372990754335535 \ep^4
%\nn&&{}-69.284610934302190975828521108391377390965135614718 \ep^5
%\nn&&{}-309.92269812465548681333650193152627910134073947626 \ep^6
%\nn&&{}-282.20149995533669796248286715145915019176339781339 \ep^7
%\nn&&{}-1248.0066201841401945462994281428236330313425072319 \ep^8
%\nn&&{}-1135.1181461481940058895700533058577982897703922890 \ep^9
%\nn&&{}-5002.0901544252559571469457974835657976170386300929 \ep^{10}
%       +\order{\e^{11}}\\
\threeBc &=&{} \label{Bc}
-0.083333333333333333333333333333333333333333333333333\ep^{-2}
\nN&&{}-0.37500000000000000000000000000000000000000000000000\ep^{-1}
\nn&&{}-2.4683003667574465515695409166563459279428082839367
\nn&&{}-8.5848042311088475775631523236940150167718153674315\ep
\nn&&{}-38.120827450450135424466436253406610052456582985006\ep^2
       +\order{\e^{3}}\\
%\nn&&{}-117.16415222705372887006149374883225459724949506810\ep^3
%\nn&&{}-444.87189813150824034237495480738674739553731856707\ep^4
%\nn&&{}-1284.4884921516316124416266660124499211124568260603\ep^5
%\nn&&{}-4537.1043939448648772411458308908236112753513002588\ep^6
%\nn&&{}-12694.960050448489830126078419463947846990909348990\ep^7
%\nn&&{}-43326.193602244506289782632936473266027694071776090\ep^8
%\nn&&{}-119373.75550899768792539088974076618505605617102152\ep^9
%\nn&&{}-400837.69395751854783487831861915269390861588188557\ep^{10}
%       +\order{\e^{11}}\\
\threeBa &=&{} \label{Ba}
+0.33333333333333333333333333333333333333333333333333\ep^{-3}
\nN&&{}+0.16666666666666666666666666666666666666666666666667\ep^{-2}
\nn&&{}+0.58333333333333333333333333333333333333333333333333\ep^{-1}
\nn&&{}+0.41381840842558476106596843069719997537329677957466
\nn&&{}-24.905969600320865917659060143145414845610363033237\ep
\nn&&{}-12.059724940640299353325034075589267393005352211165\ep^2
       +\order{\e^{3}}\\
%\nn&&{}-409.81938359563270948401137808124258058435182566228\ep^3
%\nn&&{}-213.48784390556017256884346499080700852987004498120\ep^4
%\nn&&{}-4600.7634258376003744935159553604758510892591809845\ep^5
%\nn&&{}-2463.2598612322869521123460233528191760745178766444\ep^6
%\nn&&{}-45463.000489916039633624997835362047470628224252593\ep^7
%\nn&&{}-24654.766659350671001287608884026412210216689450222\ep^8
%\nn&&{}-426328.96162455156724067604895740942754720952776295\ep^9
%\nn&&{}-232612.02621026582225827194834061760575393490058314\ep^{10}
%       +\order{\e^{11}}\\
\threeVe &=&{} \label{Ve}
-0.33333333333333333333333333333333333333333333333333\ep^{-3}
\nN&&{}-0.66666666666666666666666666666666666666666666666667\ep^{-2}
\nn&&{}-5.9565348003631195396114969999587170451045664690803\ep^{-1}
\nn&&{}-10.976993729846780032023343117902167435855817881707
\nn&&{}-67.587197404302297868575437012376235190940093056288\ep
\nn&&{}-120.04368176952781664333978568556636327245719008030\ep^{2}
       +\order{\e^{3}}\\
%\nn&&{}-656.92864363736869089291061426640171200253767348193\ep^{3}
%\nn&&{}-1149.8207535031658095428532637366713420848832732761\ep^{4}
%\nn&&{}-6054.6032800391930129998656011955165557503099447480\ep^{5}
%\nn&&{}-10543.489197355129979232548210835725725508547317700\ep^{6}
%\nn&&{}-54866.171599427440209812550919973100586674703108595\ep^{7}
%\nn&&{}-95392.648831637656093385727007512764732903369426259\ep^{8}
%\nn&&{}-494721.86436529429764810103885419358359752704324098\ep^{9}
%\nn&&{}-859753.57129080552132428897138406646778145271812507\ep^{10}
%       +\order{\e^{11}}\\
\fourBBa  &=&{} \label{BBa}
+0.08333333333333333333333333333333333333333\ep^{-3}
\nN&&{}+0.2361111111111111111111111111111111111111 \ep^{-2}
\nn&&{}+0.4189814814814814814814814814814814814815 \ep^{-1}
\nn&&{}+0.5870437170675600697079437393391752840153
\nn&&{}-38.15649063807203021274518890590965524153 \ep
       +\order{\e^{2}}\\
%\nn&&{}-109.3158094488543181647978805512546869165 \ep^2
%\nn&&{}-1354.220801557133693027078944391942499489 \ep^3
%\nn&&{}-3558.613543206844723189175735818487248061 \ep^4
%\nn&&{}-30581.00750580894139847773688637735033548 \ep^5
%\nn&&{}-77734.90703201406969583780546445895672103 \ep^6
%\nn&&{}-581483.8320640711304427359838377408167097 \ep^7
%\nn&&{}-1454201.794874518656080375874698052819711 \ep^8
%\nn&&{}-10190959.82972248783364833332679557919329 \ep^9
%\nn&&{}-25270000.17019204106491882441094937351318 \ep^{10}
%       +\order{\e^{11}}\\
\fourTa &=&{} \label{Ta}
+0.1666666666666666666666666666666666666667\ep^{-4}
\nN&&{}+0.1666666666666666666666666666666666666667\ep^{-3}
\nn&&{}-0.1666666666666666666666666666666666666667\ep^{-2}
\nn&&{}+0.1130270533440971408650083562391832748449\ep^{-1}
\nn&&{}-78.06045784302350379977045250776444270422
\nn&&{}-77.68037792722049725359022296531379645497\ep
       +\order{\e^{2}}
%\nn&&{}-2241.943000469283532268265568776842000608\ep^{2}
%\nn&&{}-2372.102274478740264968131179197892031279\ep^{3}
%\nn&&{}-46368.67128895864079729603535445835676671\ep^{4}
%\nn&&{}-50438.08680690087955066789434347397440641\ep^{5}
%\nn&&{}-843292.1685425966784939137551080611327337\ep^{6}
%\nn&&{}-930678.7486742830889694090641136686546180\ep^{7}
%\nn&&{}-14434199.88184085255757553396565199884461\ep^{8}
%\nn&&{}-16054697.85282524554858227153584899308017\ep^{9}
%\nn&&{}-239539243.1476296571163198482824927926115\ep^{10}
%       +\order{\e^{11}}
\ea

In fact, the results shown here have been obtained via
numerically evaluating truncated factorial series 
along the lines of Sections~\ref{se:two} and \ref{se:three},
but they of course coincide perfectly with the analytical results
of Section~\ref{se:Analytic} (note the different normalization there).

\end{appendix}

%%%%%%%%%%%%%%%%%%%%%%%%%%%%%%%%%%%%%%%%%%%%%%%%%%%%%%%%%%%%
%%%%%%%%%%%%%%%%%%%%%%%%%%%%%%%%%%%%%%%%%%%%%%%%%%%%%%%%%%%%
%%%%%%%%%%%%%%%%%%%%%%%%%%%%%%%%%%%%%%%%%%%%%%%%%%%%%%%%%%%%

%%%%%%%%%%%%%%%%%%%%%%%%%%%%%%%%%%%%%%%%%%%%%%%%%%%%%%%%%%%%
%%%%%%%%%%%%%%%%%%%%%%%%%%%%%%%%%%%%%%%%%%%%%%%%%%%%%%%%%%%%
%%%%%%%%%%%%%%%%%%%%%%%%%%%%%%%%%%%%%%%%%%%%%%%%%%%%%%%%%%%%

\end{document}